\documentclass[amsmath,amssymb]{revtex4}
\usepackage{graphicx}

\begin{document}

\def\beqa{\begin{eqnarray}}
\def\eeqa{\end{eqnarray}}
\def\beqn{\begin{equation}}
\def\eeqn{\end{equation}}
\def\eqa#1{\begin{eqnarray*}\mathnormal{ #1}\end{eqnarray*}}
\def\eqn#1{\begin{equation*} \mathnormal{#1}\end{equation*}}
\def\stand#1{\left[#1\right]_\mathrm{st}}

\title{Mass, inertia and gravitation}
\author{Marc-Thierry Jaekel}
\affiliation{Laboratoire de Physique Th\'eorique de l'ENS, 24 rue Lhomond, F75231 Paris
Cedex 05}
\thanks{CNRS, Ecole Normale Sup\'{e}rieure (ENS), Universit\'{e} Pierre et
Marie Curie (UPMC)}
\email{Marc.Jaekel@lpt.ens.fr}
\author{Serge Reynaud}
\affiliation{Laboratoire Kastler Brossel, case 74, Campus Jussieu, F75252 Paris Cedex 05}
\thanks{CNRS, ENS, UPMC}
\email{Serge.Reynaud@spectro.jussieu.fr}

\begin{abstract}
We discuss some effects induced by quantum field fluctuations  on mass, inertia and gravitation.
Recalling the problem raised by vacuum field fluctuations with respect to inertia and gravitation, we 
show that vacuum energy differences, such as Casimir energy, do contribute to inertia. 
Mass behaves as a quantum observable and in particular possesses quantum fluctuations. 
We show that the compatibility of the quantum nature of mass with gravitation
 can be ensured by conformal symmetries, which allow one to formulate a quantum version of the equivalence principle.
Finally, we consider some corrections to the coupling between metric fields and energy-momentum tensors 
induced by radiative corrections.
Newton gravitation constant is replaced by two different running coupling constants in the sectors of
traceless and traced tensors. There result metric extensions of general relativity,
which can be characterized by modified Ricci curvatures or by two gravitation potentials. 
The corresponding phenomenological framework   extends the usual Parametrized Post-Newtonian one,
with the ability to remain compatible with classical tests of gravity while accounting  for new features, 
such as Pioneer like anomalies or anomalous  light deflection.

\end{abstract}

\maketitle

\section{Introduction}

The relativistic conception of motion in space-time introduced by Einstein 
leads to consider inertial mass as a general property shared by all forms of energy \cite{Einstein06,Einstein07}.
The law of inertial motion reflects the underlying symmetries of space-time \cite{Einstein05b}.
Simultaneously to the development of relativity theory, Einstein laid the basis of a description of Brownian motion, {\it i.e.} motion in a fluctuating environment \cite{Einstein05a}. A fundamental role is played by fluctuation-dissipation relations, which allow one to derive the force felt by a moving body from the fluctuating force exerted by its environment. Fluctuation-dissipation relations have been shown to admit a quantum extension  \cite{Callen51,Kubo66}, which is well captured by the formalism of linear response theory \cite{Landau84}. It also appears that the vacuum state of quantum fields \cite{Planck11} is just a particular case of a thermal equilibrium state, in the limit of zero temperature. Motion in vacuum can either be considered from a relativistic point of view, using the space-time symmetries of empty space, or from a quantum point of view, using the response properties of vacuum with respect to motion. 
Then, the question naturally arises of testing the compatibility between these different approaches and of the
consequences for the notion of inertia. 

When treated in a naive way, the vacuum energy of quantum fields leads to difficulties at the cosmological level, due to the incompatibility 
of a very large value of vacuum energy with gravitational laws according to General Relativity (GR).
In this framework, the vacuum energy density is associated  with the cosmological constant and
 observational evidence pleads for a small value of the latter 
\cite{Weinberg00,Turner01}. Astrophysical and cosmological observations suggest that, at least 
within GR,  the energy of vacuum fields should not be treated as other forms of energy.
A most radical position in this respect has been
early advocated by Pauli \cite{Pauli33} (see \cite{Enz74} for an english translation):
"At this point it should be noted that it is more consistent here, in contrast to the
material oscillator, not to introduce a zero-point energy of ${1\over 2}h\nu$ per degree
of freedom. For, on one hand, the latter would give rise to an infinitely large energy per
unit volume due to the infinite number of degrees of freedom, on the other hand, it would
be principally unobservable since nor can it be emitted, absorbed or scattered and hence,
cannot be contained within walls and, as is evident from experience, neither does it
produce any gravitational field". 
This position follows the remark that since all 
types of interactions, excepting gravitation, only involve differences of energy, 
 the latter can be evaluated with respect to a minimum, for instance the vacuum, a prescription realized by normal ordering.
 Despite the fact that such a position cannot be sustained for gravitation,
as no unambiguous frame independent 
definition of the vacuum state is available in present quantum field
theories \cite{Birrell82,Fulling89}, it also appears to be wrong for other types of interaction,
as we show in the following.

Vacuum fluctuations of the electromagnetic field are known to have 
observable consequences on atoms or microscopic scatterers \cite{Cohen88,Itzykson85}.
Spontaneous emission of atoms and Lamb shifts of frequencies have been accurately measured,
in good agreement with theoretical predictions.
Van der Waals forces, which result from 
scattering of electromagnetic field vacuum fluctuations by atoms, play a crucial role
in physico-chemical processes and Casimir forces between two macroscopic electromagnetic plates, or mirrors,
due to vacuum field fluctuations \cite{Casimir48} have now been measured 
with good accuracy \cite{Bordag01}. We show below that a consistent treatment of quantum field fluctuations and of the relativistic motion 
of a scatterer entails that 
the energy of vacuum fluctuations must be taken into account in inertial effects. 
This result is obtained within the linear response formalism and follows from a consistent treatment of motion
and of quantum field fluctuations \cite{JR93b,JR97b,JR02}.

Compatibility may be maintained between  quantum and relativity theories, with the consequence 
of promoting mass to the status of a quantum observable. 
The quantum nature of the mass observable comes with a similar property for the observables describing positions in space and time.
This also conflicts with the status given to space-time positions in Quantum Field Theory (QFT).
The relativistic generalization of quantum theory has led to attribute to spatial positions the same representation as to time, 
{\it i.e.} to represent space-time positions as underlying parameters.
This position has also been inforced by arguments stating the inconsistency of defining a time operator \cite{Pauli33,Wightman62}.
We shall also show that these arguments can be bypassed and that 
the relativistic requirements of dealing with observables and the quantum representation of the latter as operators 
can both be satisfied \cite{JR96b,JR07b}. As this property also ensures compatibility with the
equivalence principle \cite{JR97a}, this revives the question of the effects of quantum fluctuations on gravitation. 

As previously remarked, GR, although in remarkable agreement with all tests of gravitation which have been performed,
is challenged by observations at galactic and cosmological scales. Besides difficulties with the cosmological constant,
anomalies  affect the rotation curves of galaxies \cite{Aguirre01,Riess98} and, as observed more recently, the relation
between redshifts and luminosities for type II supernovae \cite{Perlmutter99}.
These anomalies may be accounted for by keeping the theory of gravitation unchanged and by introducing unseen components 
under the form of dark matter and dark energy. As long as dark components remain unobserved by direct means, they 
 are equivalent to deviations from GR occuring at large scales \cite{Sanders02,Lue04,Turner04}.
In this context, the anomaly which has been observed on  Doppler tracking data registered on the Pioneer 10/11 probes \cite{Anderson98},
during their travel to the outer parts of the solar system,
constitutes a further element of questioning.

Besides its geometric setting, gravitation may be treated
within the framework of QFT as other fundamental interactions \cite{Thirring61,Feynman63,Weinberg65}.
Then, its coupling to energy-momentum tensors leads to modifications which are induced by quantum field fluctuations \cite{deWitt62,Deser74,Capper74}.
These effects, or radiative corrections, induced on gravitation by quantum fluctuations of energy-momentum tensors
are similar to those induced on motion and  may similarly be treated within the linear response formalism \cite{JR95b}.
As expected, these modifications affect the nature of gravitation at small length scales \cite{Stelle77,Sakharov67,Adler82},
but they may also affect its behavior at large length scales \cite{Nieto,Deffayet02}.
GR must then be considered as an effective theory of gravity valid at the length 
scales for which it has been accurately tested \cite{Will06} but not necessarily at other scales where deviations may occur \cite{Reuter02}.
The modified theory, while remaining in the vicinity of GR, entails deviations which may have already been observed,
and may be responsible for the Pioneer anomaly \cite{JR05a}.
This anomaly, which has escaped up to now all attempts of explanation based on the probes themselves or their spatial environment
\cite{Nieto05} may point at an anomalous behavior of gravity already occuring at scales 
of the order of the size of the solar system.
As these scales cover the domain where GR has been tested most accurately, the confrontation of the Pioneer 
anomaly with classical tests of gravitation provides a favorable opportunity for constraining the possible extensions of gravitation theory \cite{JR07a}.

This article contains three main parts. In a first section, we review how inertial motion can be treated in a way which remains consistent both with the relativistic and quantum frameworks. We obtain that vacuum energy differences, such as Casimir energy, do contribute to inertia.  
We discuss in a second section the quantum properties of mass, in particular its  fluctuations, induced by quantum field fluctuations, and its representation as an operator,  belonging to an algebra which contains both quantum observables and the generators of space-time symmetries. Finally, in a third section, we review the main features of metric extensions of GR which result from radiative corrections. We show that they lead to an  extended framework which has the ability to remain compatible with present gravity tests while accounting for Pioneer-like anomalies and predicting other related anomalies which could be tested.

\section{Vacuum fluctuations and inertia}

Most sources of conflicts between QFT and GR 
can be traced back to infinities arising when considering the contributions of
quantum fields in vacuum \cite{Planck11}.
 As a result, although vacuum fields
are responsible for now well established mechanical effects, such as Casimir forces  \cite{Casimir48}, the infinite value of vacuum energy has led to question 
its contribution to relativistic effects associated with the energy, including inertia and gravitation \cite{Pauli33}.
One usually admits that only energy differences are observable, which justifies to ignore the contribution of vacuum fields to energy, using a normal ordering prescription for all operators built on quantum fields, such as energy-momentum tensors.
However, this prescription hardly applies to gravitation, due to non linearities of GR and to the impossibilty to define vacuum in covariant way \cite{Birrell82,Fulling89}. Observations of the Universe at very large scales also plead for considering the role played by vacuum energy \cite{Turner01}.
In this section, we show that the energy corresponding to Casimir forces does indeed contribute to inertia, in full agreement with the relativistic law of inertia of energy \cite{Einstein07}. There results  that one cannot ignore the contribution of vacuum energy to the inertial effects associated with quantum fields. Furthermore, mass is affected by intrinsic quantum fluctuations and
cannot be considered any more as a mere constant parameter, but should be represented as a quantum observable, {\it i.e.}
by an operator.

\subsection{Linear response formalism}

The appropriate formalism for dealing with the effects of vacuum fluctuations on inertia is available in QFT. A physical system in space-time can be considered in a general way as a scatterer of quantum fields. Scattering is completely determined by the interaction part of the Lagarangian or Hamiltonian of the whole system, scatterer plus fields. 
A localized system may also be represented as a set of boundary conditions
for fields propagating in space-time \cite{Fulling76}, but this simplified description appears to be too crude and 
 a limiting case of the general description. Space-time evolution of the whole system is best described in the interaction picture
 \cite{Itzykson85,Barton63}, where a unitary evolution operator provides the required quantum description of all observables
 and the large time limit gives the scattering matrix for fields. 
All space-time properties of the scatterer, and in particular its motion, may be obtained from general transformations of  the interaction part of the Lagrangian or Hamiltonian.

To be more explicit, a physical system in space-time may be seen as performing a transformation of incoming free fields $\Phi^{in}$  into interacting fields $\Phi$, which at large time become outcoming free fields  again, {\it i. e.} scattered fields $\Phi^{out}$. Interacting fields $\Phi$ evolve according to a total Hamiltonian $H$ which includes a free part $H_0$ and the interaction part $H_I$. The evolution in time $t$ of the interacting fields $\Phi(t)$ is then described as an operator $S_t$ acting on incoming fields $\Phi^{in}(t)$. Free fields evolve according to the free part $H_0$ of the Hamiltonian ($\hbar$ is Planck constant, $[, ]$ denotes the commutator of two operators)
\beqa
\label{time_evolution}
&&{d \Phi\over dt}= -{i\over\hbar}[H, \Phi], \qquad H \equiv H(\Phi)\nonumber\\
&&{d \Phi^{in}\over dt}= -{i\over\hbar}[H_0^{in}, \Phi^{in}], \qquad H_0 ^{in}\equiv H_0(\Phi^{in})\nonumber\\
&&\Phi(t) \equiv S_t^{-1} \Phi^{in}(t) S_t
\eeqa
The evolution operator $S_t$ is obtained from the  interaction part $H_I$ of the Hamiltonian by solving a differential equation resulting from the evolution of interacting and free  fields (\ref{time_evolution})
\cite{Itzykson85}
\beqa
\label{evolution_operator}
&&{d S_t \over dt}= {i\over\hbar}H_I^{in}(t) S_t, \quad H \equiv H_0 + H_I, \qquad  H_I^{in}\equiv H_I(\Phi^{in}) \nonumber\\
&&S_t = {\cal T} exp {i\over \hbar} \int_{-\infty}^t H_I^{in}(t') dt'
\eeqa
${\cal T}$ denotes time ordering for the different terms building the exponential, while
$H^{in}_I$ denotes the interaction Hamiltonian written in terms of input fields $\Phi^{in}$ ($H_I(t)= S_t^{-1} H^{in}_It) S_t$ is the same expression, but written in terms of interacting fields $\Phi$). Equations (\ref{evolution_operator}) determine the interacting fields in terms of input fields only. Similarly, output fields are determined by a scattering matrix $S$ which is the large time limit of the evolution operator $S_t$
\beqa
\label{S-matrix}
\Phi^{out}(t) = S^{-1} \Phi^{in}(t) S, \qquad S = S_\infty
\eeqa
This description applies to a scatterer at rest or in motion indifferently, the only difference  between the two cases lying in a modification of the interaction Hamiltonian describing the scatterer. 
All properties of the scatterer with respect to its localization or motion in space-time are encoded in the interaction term. 

Motions correspond to modifications of the interaction Hamiltonian and, as long as they can be considered as small perturbations, may be treated using the formalism of linear response theory \cite{Kubo66}.  
As can be seen from evolution equations (\ref{time_evolution},\ref{evolution_operator}),  a small perturbation of the interaction Hamiltonian (proportional to a time dependent parameter $\delta \lambda(t)$) induces a perturbation of the evolution operator, hence of all observables which are built on interacting fields:
\beqa
\label{linear_response}
&&S_t^{-1} \delta S_t= {i\over \hbar} \int_{-\infty}^t \delta H_I(t') dt', \qquad S_t^{-1}\delta H_I^{in}(t) S_t = \delta H_I(t) \equiv  B(t)\delta \lambda(t)\nonumber\\
&&\delta A(t) = [A(t), S_t^{-1} \delta S_t] =  \int_{-\infty}^\infty {i \over \hbar}\theta (t-t') [A(t),B(t')]\delta \lambda(t') dt'\nonumber\\
&& \qquad \theta(t) \equiv 0 \quad {\rm{for}} \quad t < 0, \quad \theta(t) \equiv 1 \quad {\rm{for}} \quad t > 0
\nonumber\\
&&<\delta A(t)> = \int_{-\infty}^\infty  \chi_{AB}(t,t') \delta \lambda(t')dt'\nonumber\\
&&\qquad \chi_{AB}(t,t') = {i \over \hbar}\theta (t-t') <[A(t),B(t')]>
\eeqa
The linear response of an observable $A$ to a perturbation may  be written as the action of a generator $B$ and the response is captured in a susceptibility function $\chi_{AB}$, which only depends on the commutator of $B$ with $A$.
Evolution equations (\ref{evolution_operator}) and hence the general form (\ref{linear_response}) of linear response hold independently of the particular form taken by the interaction Hamiltonian.
Causality of responses follows from the time ordering prescription entering the definition of the evolution operator (\ref{evolution_operator}), as is made explicit in linear responses (\ref{linear_response}) by the presence of Heaviside step function  $\theta$ in the time domain. When response functions are written in the frequency domain (after a Fourier transformation), causality equivalently corresponds to  analyticity in the upper half of the complex plane.
Kramers-Kronig relations then allow one to relate the real and imaginary parts of the susceptibility function
\beqa
\label{KK-relations}
&&\chi(t)\equiv  \int_{-\infty}^\infty {d\omega\over2\pi} e^{-i\omega t}\chi[\omega], \qquad \chi\equiv Re(\chi)+ i Im(\chi)\nonumber\\
&&\chi[\omega] = \int_{-\infty}^\infty {d\omega^\prime\over\pi}{Im(\chi)[\omega^\prime]\over\omega^\prime-\omega -i\varepsilon}
\eeqa 
For simplicity, invariance under time translation has been assumed; $\varepsilon\rightarrow 0^+$ defines the integration contour in the frequency complex plane.

In general, as can be seen on explicit cases dealing with quantum fields, short time singularities occur which generate infinities in time ordered products (\ref{evolution_operator}). These infinities take their origin in divergences occuring at high frequencies and  are well-known in QFT where they are  treated by renormalization. Renormalization amounts to modify the interaction with additional counterterms to compensate the divergencies and by imposing renormalization conditions on
final observables to fix the resulting ambiguities \cite{Itzykson85}. These counterterms modify the $\theta(t-t^\prime)$ factor defining time ordered products  (they are localized at equal times $t=t^\prime$)  and affect the reactive part of the susceptibility. They amount to modify  dispersion relations (\ref{KK-relations}) by subtractions (of a finite number of terms of a Laurent expansion of the response function) and the resulting ambiguities are raised with the renormalization prescriptions. 
Equation (\ref{linear_response}) then correspond to unambiguous relations, in the frequency domain,
only between the dissipative part $Im(\chi_{AB})$ of the susceptibility and the commutator $\xi_{AB}$
\beqa
\label{fluctuations_dissipation}
&&\xi_{AB}(t-t^\prime) \equiv {1\over2\hbar}<[A(t), B(t^\prime)]> \equiv \int_{-\infty}^\infty {d\omega\over2\pi} e^{-i\omega (t-t^\prime)}\xi_{AB}[\omega]\nonumber\\
&&Im(\chi_{AB})[\omega] = \xi_{AB}[\omega]
\eeqa
Determining the reactive part $Re(\chi_{AB})$ of the susceptibility requires either to compute the susceptibility or to use dispersion relations (\ref{KK-relations}) with subtractions and additional constraints. Both ways must treat the same infinities and the additional constraints are provided by the prescriptions accompanying renormalization. Renormalization allows one to treat infinities at the expense of a detailed treatment of the interaction, at least at a perturbative level. In the following, we shall focus on characterizing responses in the most general way as possible, without entering a detailed description
of interactions, and shall let aside the questions raised by criteria to be fullfilled for renormalizability. 
 
Equations (\ref{linear_response},\ref{fluctuations_dissipation}) show that the response of a system to a perturbation is strongly constrained  by the quantum correlations of observables describing the unperturbed system. When the latter is in a vacuum state, quantum correlations may be further characterized, as vacuum can  be seen as a thermal state at the limit of 
zero temperature. For a general thermal equilibrium, namely for a state described by a thermal density matrix,
quantum correlations satisfy Kubo-Martin-Schwinger (KMS) relations \cite{Callen51,Kubo66,Itzykson85}
($T$ is the temperature, Boltzman constant is set equal to $1$, Tr is the trace in the Hilbert space providing mean values for observables)
\beqa
\label{KMS_relations}
&& C_{AB}(t-t^\prime)\equiv <A(t) B(t^\prime)> - <A(t)>< B(t^\prime)>\nonumber\\
&& \qquad <A> \equiv  {\rm{Tr}}\left(\rho A\right), \qquad \rho = {e^{-{\hbar H \over T}} \over{\rm{Tr}} \left(e^{-{\hbar H \over T}}\right)}\nonumber\\
&&2\hbar \xi_{AB}(t) \equiv C_{AB}(t) - C_{BA}(-t), \qquad \qquad 2\hbar \sigma_{AB}(t) \equiv C_{AB}(t) + C_{BA}(-t)
\nonumber\\
&&2\hbar \xi_{AB}[\omega]
= (1 - e^{-{\hbar \omega \over T}}) C_{AB}[\omega], \qquad \qquad
\sigma_{AB}[\omega] = {\rm{coth}} ({\hbar\omega \over 2T}) \xi_{AB}[\omega]
\eeqa
In the zero temperature limit, these relations characterize the vacuum state and 
allow one to deduce all correlation functions from commutators (${\rm{sgn}}$ is the
sign function)
\beqn
\label{KMS_vacuum}
C_{AB}[\omega] = 2\hbar \theta(\omega) \xi_{AB}[\omega], \qquad \qquad
\sigma_{AB}[\omega] = {\rm{sgn}}(\omega) \xi_{AB}[\omega]
\eeqn
As expected for fluctuations in the ground state, correlations are stationary, with a spectrum limited to positive frequencies only.  In particular, this entails that the commutator in the time domain cannot vanish, showing the irreducible non commuting character of quantum fluctuations. In the case of correlations of free quantum fields, commutators become pure numbers and mean values may be omitted
\beqa
\label{free_field_commutator}
&&2\hbar\xi_{\Phi_{in}(x)\Phi_{in}( x^\prime)}(t-t^\prime) = [\Phi_{in}(t,x), \Phi_{in}(t^\prime,x^\prime)]\nonumber\\
&&\chi_{\Phi\Phi}(t-t^\prime, x- x^\prime ) = 2i\theta(t-t^\prime)\xi_{\Phi_{in}(x)\Phi_{in}(x^\prime)}(t-t^\prime)
\eeqa
The response of a quantum field to its source, {\it i.e} the retarded propagator, is given by the  correlation function associated with the field commutator.

\subsection{Response to motions}

In this part, we shall only consider physical systems evolving in a flat space-time.
Hence, displacements of these systems, including quantum fields, may be associated with  symmetries of space-time. According to Noether theorem, when evolution, {\it i.e.} propagation of fields and interaction, satisfies some symmetries, the corresponding  generators are associated with conserved quantities \cite{Itzykson85}.
Invariance under space-time translations implies conservation of energy-momentum, which corresponds to the generator of translation symmetry.
In quantum theory, the generators of space-time symmetries acting on quantum observables and the corresponding conserved quantities  are built from the energy-momentum tensor. In particular, displacements $\delta q$  in space  are generated by the operator corresponding to momentum $P$. 

For a system made of quantum fields and a scatterer coupled to them, the generator of space translation for the whole system identifies with the total momentum, which is conserved when the coupling between fields and scatterer is a scalar, {\it i.e.} a space-time invariant.
This property may be stated in an equivalent way as the invariance under translation of the interaction part in the total Lagrangian or Hamiltonian.
Hence, the momenta of field and scatterer, which generate their respective translations 
in space, have opposite actions. The action of a translation of the scatterer on its coupling to fields is equivalent to the action of the opposite of the field momentum.
Then, according to the previous section, perturbation of the interaction Hamiltonian $H_I$  under displacements of the scatterer is provided by the commutator of the field momentum $P$ with $H_I$ (\ref{linear_response}) 
\beqa
\label{displacements}
\delta H_I  = -{i \over \hbar}[H_I, P]\delta q= -{i \over \hbar}[H, P]\delta q  = F \delta q, \quad F\equiv {d P\over dt} = -{i \over \hbar}[H_I, P]
\eeqa
Free field propagation being invariant under space-time translation, the field momentum $P$ is conserved during propagation, hence commutes with $H_0$.
As shown by equations (\ref{displacements}), small motions $\mathnormal{\delta q}$ of the scatterer and their effect on scattered quantum fields (\ref{linear_response})  are completely determined by a single observable $F$, the force felt by the scatterer.
 Then,  perturbations of observables due to the scatterer's motions are obtained from
 their commutator with 
 the force, {\it i.e.} the radiation pressure, exerted by scattered fields.

In particular, when a scatterer is set into motion, the force exerted by scattered fields undergoes itself a perturbation $\delta F$ which is related to the scatterer's displacement $\delta q$. According to the linear reponse equations (\ref{linear_response},\ref{fluctuations_dissipation}) and the general form (\ref{displacements}) of the generator of displacements, this relation may be written as a motional susceptibility $\chi_{FF}$ which is determined by fluctuations of the force $\xi_{FF}$, exerted on the scatterer at rest
\beqa
\label{force_response}
&&<\delta F(t)> =  \int_{-\infty}^\infty\chi_{FF}(t-t') \delta q (t') dt' \nonumber\\
&&\chi_{FF}(t) =2i\theta (t)\xi_{FF} (t), \qquad
\xi_{FF} (t-t^\prime) ={1\over2\hbar}<\left[ F(t),F(t^\prime)\right]>
\eeqa
A warning remark must be made at this point. As underlined in previous section when discussing the causality properties of response functions (\ref{KK-relations}), the relation (\ref{force_response}) between motional susceptibilities and force correlations in fact suffers from ambiguities due to necessary subtractions in Kramers-Kronig relations. These are related to ambiguities affecting the time ordering defining the evolution operator (\ref{evolution_operator})
 at the limit of equal times, due to a necessary treatment of divergences. The additional constraints allowing one to raise 
these ambiguities are equivalent to the
 renormalization prescriptions  following the regularization of infinities.
The same ambiguities affect the definition (\ref{displacements}) of the generator of motional perturbations, built in terms of  generators of space-time symmetries.
Then, a complete and self-consistent treatment of motional perturbations would require a discussion
in terms of renormalized quantities, including in particular the generator of motional perturbations, {\it i.e} the force $F$. As we shall not enter such a discussion in the present article, we must keep in mind that some defects might arise, which are due to an incomplete description and which can be cured by a more detailed treatment
involving renormalization of physical quantities.

The force induced by quantum fields on a moving scatterer does not depend on the type of coupling describing the interaction between scatterer and fields, but only on the balance of energy-momentum resulting from overall energy-momentum conservation. 
This general property may be seen as a consequence of the fundamental connection between motions in space-time and symmetries which underlies relativity theory. 
Response functions to motion are then determined by correlation functions of the energy-momentum tensor of scattered quantum fields.  
One then expects that the description of forces induced by motion in vacuum fields remains consistent with the general principles of relativity theory
and in particular with the law of inertia of energy \cite{Einstein06}.
In the following, we show that this law is indeed respected by vacuum energies, by studying exemplary cases built with mirrors and cavities.
 
For the sake of simplicity, and in order to discuss explicit expressions, we shall illustrate the following arguments on a simplified description of a scatterer \cite{JR92a}. Without essential reduction of generality, we shall consider quantum fields in a two-dimensional space-time, {\it i.e.} fields propagating in a single spatial dimension ($\varphi(t-x)$ and $\psi(t+x)$ for the two different directions)
\beqa
\label{2d-fields}
&&\phi(t,x) \equiv \varphi(t-x) + \psi(t+x), \qquad \Phi[\omega] \equiv \left[ \begin{array}{c} \varphi[\omega] \\ \psi[\omega]\end{array} \right]\nonumber\\
&&\xi_{\Phi_{in}\Phi_{in}}[\omega] =\left[\begin{array}{cc}\xi_{\varphi_{in}\varphi_{in}}[\omega]&0\\0& \xi_{\psi_{in}\psi_{in}}[\omega]\end{array}\right] ={1 \over\omega}
\eeqa
The scatterer will also be represented  by a scattering matrix $S$ mixing the two counterpropagating directions and corresponding to a scatterer localized at a spatial position $q$. Considering  scatterers with a mass which is large when compared with the field energy, we shall also neglect recoil effects, so that the scattering matrix (\ref{S-matrix}) becomes a quadratic form of input fields and may be rewritten under the simple form of a 2x2 matrix acting on field components (written in the frequency domain)
\beqa
\label{scattering_matrix}
&&\Phi_{out}[\omega] = S[\omega] \Phi_{in}[\omega], \qquad  S[\omega] \equiv  \left[ \begin{array}{cc} s[\omega]& r[\omega]e^{-i\omega q}\\ r[\omega]e^{i\omega q}& s[\omega]\end{array} \right]
\eeqa
$r[\omega]$ and $s[\omega]$ respectively describe the frequency dependent reflection and transmission coefficients associated with field scattering, the latter being defined  in a reference frame which is comoving with the scatterer; $q$ describes the  spatial position of the scatterer in this frame.
Besides its usual analyticity and unitarity properties ($|r|^2 + |s|^2 =1$), the scattering matrix $S$ will also be assumed to approach the identity matrix above some frequency cut-off, smaller than the scatterer's mass.
This condition corresponds to transparency at high frequencies and, besides being indeed satisfied by real mirrors, allows one to neglect recoil effects. Furthermore, this property will provide a natural (physical) high frequency regulator, thus leading to finite results.

The energy-momentum tensor $(T^{\mu\nu})_{\mu,\nu=0,1}$ of propagating fields (\ref{2d-fields}) is determined from their space-time derivatives
\beqa
\label{stress_tensor}
&&T^{00} = T^{11} = {1\over2}(\partial_t \phi^2 +
\partial_x \phi^2) = \dot{\varphi}^2 + \dot{\psi}^2, \qquad \dot{\varphi}\equiv \partial_t\varphi, \quad \dot{\psi}\equiv \partial_t\psi\nonumber\\
&&T^{01} = T^{10} = - \; \partial_t \phi \;
\partial_x \phi = \dot{\varphi}^2 - \dot{\psi}^2
\eeqa
Hence, the force $F$ exerted on the scatterer is obtained from the balance of momenta of incoming and outcoming fields (of energy-momentum fluxes through the scatterer's surface)
\beqa
\label{force}
&&F(t)= \dot{\varphi}_{in}^2(t-q) - \dot{\varphi}_{out}^2(t-q) -\dot{\psi}_{in}^2(t+q) + \dot{\psi}_{out}^2(t+q)
\eeqa 
The force exerted on the scatterer in vacuum is then determined from the scattering matrix (\ref{S-matrix}, \ref{scattering_matrix}),  KMS relations (\ref{KMS_vacuum})   and commutators  of incoming free fields
(\ref{free_field_commutator}, \ref{2d-fields}).
Equations (\ref{force}) and S-matrix unitarity imply that the mean force $<F(t)>$ exerted on a scatterer at rest in vacuum vanishes. On another hand, force correlations in vacuum $\xi_{FF}(t)$ do not vanish and remain finite  due to the high frequency transparency of the scattering matrix (\ref{scattering_matrix}).

Motions of the scatterer $\delta q(t)$ ($\delta q[\omega]$ in the frequency domain) modify the field scattering matrix $S$. The S-matrix of a moving scatterer can be directly obtained by applying a frame transformation to the S-matrix at rest. The 
corresponding frame transformation on fields (\ref{2d-fields}), hence on scattering matrices (\ref{scattering_matrix}), is provided by the change of coordinates which transforms the laboratory frame into a frame which is comoving with the scatterer.  
 The transformation
 results in a modified scattering matrix $S+\delta S$  which may still be described by a 2x2 matrix, 
but with changes of frequency:
\beqa
\label{perturbed_S_matrix}
&&\delta\Phi_{out}[\omega] = 
\int_{-\infty}^\infty {d\omega'\over2\pi} \delta S[\omega,\omega']
\Phi_{in}[\omega']\nonumber\\
&&\delta S[\omega,\omega'] = i\omega^\prime\delta q[\omega-\omega^\prime]\left(S[\omega]  \left[\begin{array}{cc}1&0\\0&-1\end{array}\right] - \left[\begin{array}{cc}1&0\\0&-1\end{array}\right] S[\omega^\prime]\right)
\eeqa
As discussed in previous section, modifications of the scattering matrix (\ref{linear_response})
may also be derived from the perturbation induced by the generator associated with the force exerted on the scatterer (\ref{displacements}). One checks that the latter, a quadratic form of input free fields (\ref{force}), as the scattering matrix itself (\ref{scattering_matrix}), induces a perturbation of the scattering matrix (\ref{linear_response}) which is identical to equations (\ref{perturbed_S_matrix}).
 This property  illustrates, on this simple scattering model, the
 general connection between motions and symmetry generators entailed by the principles of
 relativity theory.

In general, vacuum input fields are transformed by the moving scatterer into output fields which are no more in the vacuum state. In other words, general motions of the scatterer lead to radiation. In return, as a consequence of energy-momentum conservation, the moving scatterer feels a mean radiation reaction force $<\delta F>$. For small perturbations, the reaction force is proportional to the scatterer's motion $\delta q$. The perturbed force may be directly obtained from expression 
(\ref{force}) and the perturbed scattering matrix (\ref{perturbed_S_matrix}), and may be expressed in terms of a motional susceptibility \cite{JR92a}
\beqa
\label{mean_force_response}
&&<\delta F[\omega]> = \chi_{FF}[\omega] \delta q[\omega]\nonumber\\
&&\chi_{FF}[\omega] = i\hbar \int_0^\omega {d\omega'\over2\pi} \omega'(\omega -
\omega')
\lbrace 1 - s[\omega']s[\omega-\omega'] +
r[\omega']r[\omega-\omega']\rbrace\nonumber\\
\eeqa
One verifies that the previously discussed relation between the dissipative part of the force susceptibility $Im(\chi_{FF})$ and force fluctuations $\xi_{FF}$ (\ref{force_response}) is satisfied by expression (\ref{mean_force_response}) \cite{JR92a}.  
One also notes that, as a result of the analyticity properties of the scattering matrix itself, expression (\ref{mean_force_response}) for the motional susceptibility 
satisfies the analyticity, {\it i.e.} causality, properties which are  characteristic of response functions. 
At perfect reflection ($r[\omega] \equiv -1, s[\omega] \equiv 0$), one also recovers from equations (\ref{mean_force_response}) the known dissipative force for
 a perfect mirror moving in vacuum,
when treated as a boundary condition for quantum fields \cite{Fulling76}
\beqa
\label{perfect_reflection}
<\delta F(t)> = {\hbar \over 6\pi} {d^3\over dt^3}\delta q(t)
\eeqa
One remarks that this result has been obtained in the present formalism without explicitly treating the infinite energy of vacuum.
This is due to the frequency dependence of the scattering matrix (\ref{scattering_matrix}), and more precisely to its high frequency transparency, the representation as a scatterer thus providing a more physical description than the treatment as boundary conditions.   
But one should remind that, as previously underlined, neither of these treatments really takes into account the fundamental divergences associated with  space-time singularities and affecting the evolution of quantum fields. These divergences require  a renormalization prescription to be managed in a self-consistent way.  
One also notes that the general expression (\ref{mean_force_response}) for the radiation reaction force in vacuum not only satisfies analyticity properties but also further positivity
properties \cite{Meixner65}, entailed by the ground state nature of the vacuum state (\ref{KMS_vacuum}).
As a result, it can be shown  \cite{JR92b} that the radiation reaction force described by equations (\ref{mean_force_response}) does not produce  unstable motions, known as "runaway solutions" \cite{Rohrlich65}, as those which would be induced by expression (\ref{perfect_reflection}).

The simple scattering model discussed here provides an explicit example of linear response formalism applied to perturbations induced by motions in space-time. 
This formalism may further be used to analyse  the Brownian motion undergone by a scatterer embedded in vacuum fields \cite{JR93a}. It allows one to obtain relations between fluctuations of positions in vacuum and the susceptibility describing the response of the system to an applied external force.

\subsection{Relativity of motion}

The mechanical effects induced on a scatterer by quantum field fluctuations allow for a complete quantum description, using the linear response formalism. Their compatibility with relativity theory
can also be checked explicitly. In particular, the energy-momentum balance responsible for the motional response of a scatterer
leads to the same result when analysed in different refrence frames.
This can be shown using the linear response formalism and the representation of motion as the action of  space-time symmetries.

The force exerted by  quantum fields on a scatterer (\ref{force}) is obtained as a balance of momentum between  outcoming and incoming fields, hence in terms of 
the scattering matrix and input field correlations only.
For the simple scattering model introduced in previous section, this reads, using expressions (\ref{stress_tensor}, \ref{force})
\beqa
\label{input_correlations}
\Phi _{out}[\omega ] =S[\omega ]\Phi _{in}[\omega ]&&\nonumber\\
<\Phi _{in}[\omega ]\Phi _{in}[\omega^\prime ]> \equiv C _{in}[\omega,\omega^\prime]&&\quad \rightarrow \quad   <F[\omega]> \equiv {\cal F}\lbrace S, C_{in}\rbrace[\omega]
\eeqa
The functional dependence  ${\cal F}\lbrace S, C_{in}\rbrace$ of the force $F$ in terms of the scattering matrix
and incoming field correlations (\ref{input_correlations}) takes a form which does not depend on the choice of a particular reference frame.
Scattering may then equivalently be analysed either in a laboratory frame, where the scatterer is moving, or in a comoving frame where the latter is at rest.  

In the laboratory frame, the scatterer's motion induces a perturbation $\delta S$ of the scattering matrix. The latter  is obtained by applying a transformation to the scattering matrix at rest (\ref{scattering_matrix}), which  corresponds either to a change of coordinates from the comoving frame to the laboratory, or to the action of a displacement generator (\ref{displacements}), both transformations providing the same result $\delta S(\delta q)$ (\ref{perturbed_S_matrix}). When analysed in the laboratory frame, incoming fields, hence their correlations $C_{in}$, remain unperturbed by the scatterer's motion (\ref{2d-fields}), so that the perturbed force reads
\beqa
\label{laboratory_force}
&&<F[\omega] + \delta F[\omega]> =   {\cal F}\lbrace S+\delta S(\delta q), C_{in}\rbrace[\omega]\nonumber\\
\rightarrow \quad &&<\delta F[\omega]> = \chi_{FF}[\omega] \delta q[\omega]
\eeqa

In a comoving frame, the scattering matrix $S$, which describes the coupling between scatterer and fields, remains unchanged (\ref{scattering_matrix}). On another hand, the space-time expression of incoming field correlations undergoes a modification $\delta C_{in}$ which  can be obtained from field correlations in the laboratory (\ref{2d-fields}) and a change of coordinates from the laboratory to a comoving frame $\delta C_{in}(\delta q)$, so that the perturbed force reads in that case
\beqa
\label{comoving_force}
&&<F[\omega] + \delta F[\omega]> =   {\cal F}\lbrace S, C_{in} + \delta C_{in}(\delta q)\rbrace[\omega]\nonumber\\
\rightarrow \quad &&<\delta F[\omega]> = \chi_{FF}[\omega] \delta q[\omega]
\eeqa
In either reference frame, the perturbation of the force exerted on the scatterer may be expressed
as a motional susceptibility $\chi_{FF}$. Both expressions (\ref{laboratory_force}) and (\ref{comoving_force}) can be seen to lead to the same result (\ref{mean_force_response}) \cite{JR92a}.
This property, here illustrated on a simple model, is in fact a general consequence of the principles ruling the evolution of quantum observables (\ref{evolution_operator}) and their space-time transformations (\ref{displacements}). Interesting applications of this property may be envisaged.
For instance, motion can be simulated by keeping the scatterer at rest but by modifying the field fluctuations reflected by the scatterer, using for instance optical devices acting on incoming fields, thus obtaining radiation and a reaction force equivalent to those induced on a scatterer by its motion \cite{JR92a}.

When combined with symmetry properties of field fluctuations, relativity of motion leads to
important consequences for the radiation reaction force. 
If the scatterer's motion corresponds to a generator of a symmetry of field fluctuations, {\it i.e.} to a frame transformation which leaves field fluctuations invariant, then the radiation reaction force is the same as for a scatterer at rest embedded in the same field fluctuations. This is in particular true for motions with uniform velocity in vacuum, due to Lorentz invariance of vacuum field fluctuations. As a result,  the mean radiation reaction force vanishes for uniform motions in vacuum. On another hand, uniform motions in a thermal bath 
induce a dissipative reaction force proportional to the scatterer's velocity, in conformity with
the transformation properties of the thermal bath under a Lorentz transformation. These properties of the radiation reaction force  may be checked by explicit computation in the previously discussed model
\cite{JR93c}.

Quantum field fluctuations in vacuum may 
be invariant under a larger group than the Poincar\'e group which is generated by translations and Lorentz  symmetries.
This is the case for the vacuum fluctuations of a scalar field in 2-dimensional space-time
or of electromagnetic fields in 4-dimensional space-time. Both admit invariance under the action of the larger group of conformal symmetries, which include generators performing transformations to uniformly accelerated frames. As a consequence, the radiation reaction force in vacuum should be invariant
under conformal transformations and uniformly accelerated motions should lead to a vanishing radiation reaction force.
Indeed, scalar fields in a 2-dimensional space-time lead to vacuum field correlations behaving as $\omega^3$ at low frequency, corresponding to a radiation reaction force behaving as the third time derivative of the position (see  (\ref{mean_force_response}))
\beqa
&&C_{FF}[\omega] \sim \theta(\omega)\omega^3, \qquad \omega \sim 0\nonumber\\
&&<\delta F> \sim {d^3\over dt^3}\delta q(t)
\eeqa
At perfect reflection, these relations hold at all frequencies, as shown by expression  (\ref{perfect_reflection}) already obtained within the formalism of quantum fields with prescribed boundary conditions \cite{Fulling76}.
 The case of a scatterer embedded in electromagnetic fields $(A_\mu)_{\mu=0,1,2,3}$
 in 4-dimensional space-time similarly involves
conformally invariant vacuum field fluctuations \cite{JR95a}, which may be written, after an appropriate gauge transformation
\beqa
C_{A_\mu A_\nu }\left( x,x^{\prime }\right) =\frac \hbar \pi
{\eta _{\mu \nu}\over\left( x-x^{\prime }\right)
^2-i\varepsilon \left( t-t^{\prime }\right) }, \qquad \eta_{\mu\nu}\equiv{\rm{diag}}(1,-1,-1,-1), \quad \varepsilon \rightarrow 0^{+} 
\nonumber\\
\eeqa
As a consequence, the radiation reaction force induced on the scatterer by its motion vanishes for uniform velocities and also for uniform accelerations ($\tau$ denotes proper time)
\beqa
<\delta F^\mu> \sim  {d^3  q^\mu\over d\tau^3} + ({d^2 q\over d\tau^2})^2 {dq^\mu\over d\tau}
\eeqa
 It can be seen that the radiation reaction force in that case is
  proportional to a conformally
 invariant derivative of the scatterer's position, {\it i.e.}  to Abraham-Lorentz vector \cite{JR95a}.

\subsection{Inertia of vacuum fields}

One should not be lead to conclude from the previous discussion that the vacuum of conformally invariant fields
does not induce any motional effect on an accelerated scatterer. 
In fact, as previously remarked, only the dissipative part (imaginary part in the frequency domain) of the motional susceptibility
is unambiguously related to correlations of quantum fields (\ref{fluctuations_dissipation}).  The  reactive part (real part in the frequency domain) of the motional susceptibility is related to the latter through Kramers-Kronig relations  (\ref{KK-relations}) but  requires more care to be determined. 
Although the general properties of the dissipative part of the motional force should remain unchanged by a renormalization of
observables built on the energy-momentum tensor of quantum fields, the properties of the reactive part in general will be sensitive to this procedure.
The model of a scatterer used in previous sections involves approximations which allow one to bypass the problems raised by coupling to high frequencies, but it appears insufficient to
examine in a general way the relation between acceleration and motional forces.
For that reason, we first extend the simple model of a single mirror to a model of a cavity, built with two mirrors which are still modelized in the same way.
This model of a cavity will allow us to exhibit a fundamental relation between vacuum fields and the response of an embedded system to accelerated motion.
Although still by-passing  fundamental space-time singularities asociated with quantum fields, 
the spatial extension introduced by a cavity will appear sufficient to exhibit a fundamental relation which was occulted by the approximations underlying the model of a single mirror. 

Still considering fields propagating in a single spatial direction (\ref{2d-fields}), each mirror building the cavity can be described by its scattering matrix ($S_i, i=1,2$), depending on its position ($q_i$) (\ref{scattering_matrix}). As in the case of a single mirror, each scattering matrix determines outcoming fields in terms of incoming ones, with the only difference that intracavity fields are also involved and play the role of incoming or outcoming fields, according to the mirror on which they reflect. The scattering matrix $S$ for the total cavity may easily be obtained from the relation made by the two mirrors' scattering matrices on overall output and input fields.
There results a relation between the determinants of the different scattering matrices which may be written
\beqa
\label{phase_shift}
&&det S[\omega ]=det S_1 [\omega ]det S_2 [\omega ]e^{2i\Delta_q [\omega ]}\nonumber\\
&&\Delta_q[\omega] = {i\over2} Log {1 - r[\omega]e^{2i\omega q/c} \over
1 - r[\omega]^* e^{-2i\omega q/c}}
\eeqa    
$q=q_2 - q_1$ denotes the spatial distance of the two mirrors, {\it i.e.} the length of the cavity, and $r[\omega]$ is the product of the frequency dependent reflection coefficients of the two mirrors.
All dependence on the spatial extension of the system is then captured in a phase-shift $\Delta_q[\omega]$. This corresponds to the phase-shift undergone by incoming fields at frequency $\omega$ when scattered by a cavity of length $q$.

When applied to  a cavity at rest, the balance of energy-momentum (\ref{stress_tensor})  between intracavity and exterior (input and output) fields leads to a difference between the two mean
forces $<F_i>, i=1,2$ acting on the two mirrors. The resulting force corresponds to the mean Casimir force $F_C$ exerted by field fluctuations between the two sides of the cavity.
When computed for input fields in vacuum, the mean Casimir force, and the corresponding Casimir energy, 
take a simple and suggestive form in terms of scattering matrices or  of  phase-shifts (\ref{phase_shift}) \cite{JR91}
\beqa
\label{Casimir_force}
&&<F_2>-<F_1> =  F_C \equiv\partial _{q}E_C\nonumber\\
&&E_C=\int_0 ^{\infty }\frac{{\rm d}\omega }{2\pi }\hbar \omega 
 \tau[\omega], \qquad \tau[\omega]\equiv {1\over2}\partial _\omega \Delta_q [\omega ]
\eeqa
The Casimir energy $E_C$ may be seen as a sum over all field modes (two counterpropagating modes for each frequency $\omega$) of a contribution built from two factors, $\hbar\omega/2$ the vacuum energy at the corresponding frequency and $\tau[\omega]$ the derivative of the phase-shift at the same frequency. The factor $\tau[\omega]$ may also be seen as the time delay undergone by a field around frequency $\omega$ during its scattering by the cavity. Then, expression (\ref{Casimir_force}) for the Casimir energy may be given a simple interpretation: it corresponds to a part of the energy of vacuum fields which is stored inside the cavity, during scattering of vacuum fields \cite{JR91}. 

As in the case of a single mirror, the forces $F_i, i=1,2$ acting on the two mirrors of a cavity
show fluctuations in vacuum which may be obtained from the mirrors' scattering matrices (\ref{scattering_matrix}) and free field correlation functions (\ref{2d-fields}), and which satisfy KMS relations in  vacuum (\ref{KMS_vacuum})
\beqa
\label{cavity_force_fluctuations}
C_{F_i F_j}(t)=\left\langle F_{i}(t)F_{j}(0)\right\rangle -\left\langle
F_{i}\right\rangle \left\langle F_{j}\right\rangle
\eeqa
Correlation functions for fluctuations of the Casimir force follow from (\ref{cavity_force_fluctuations}) \cite{JR92c}. 

Now, we consider that the two mirrors of the  cavity are moving independently, and we directly determine the corresponding motional responses.
The motions of the two mirrors correspond to perturbations (\ref{displacements}) of  the Hamiltonian
coupling the mirrors to scattered fields
\beqa
\delta H_I(t)=\sum_{i}F_{i}(t)\delta q_{i}(t)
\eeqa
Output and intracavity fields are then modified according to the perturbed scattering matrices 
(\ref{perturbed_S_matrix}) and the resulting perturbations $<\delta F_i>$ of radiation pressure on the mirrors can be expressed at first order in the mirrors' displacements $\delta q_i$ under the form of 
susceptibilities $\chi_{F_i F_j}$
\beqa
&&<\delta F_i[\omega]> = \sum_j \chi_{F_iF_j}[\omega] \delta q_j[\omega]
\eeqa
Each mirror's motion induces a motional force on both mirrors and motional susceptibilities can be checked to satisfy fluctuation-dissipation relations \cite{JR92c} with  Casimir force fluctuations 
\beqa
&&\chi_{F_iF_j}[\omega] - \chi_{F_jF_i}[-\omega] =
{i \over \hbar}\lbrace C_{F_iF_j}[\omega] -  C_{F_jF_i}[-\omega]\rbrace
\eeqa
As a result of their dependence on the energy-density of intracavity fields, motional forces can be seen to be resonantly enhanced
for motions at proper frequencies of the cavity ($\omega_n = n {\pi \over q}$).
This resonance property has important consequences, as in particular it allows one to envisage  
experimental setups for exhibiting the motional effects of vacuum fields.  Indeed, the
relatively small magnitude of these effects in vacuum,  as suggested by Casimir
forces, can be compensated by the cavity finesse and lead to observable effects \cite{JR96}.

We now focus on global motions of the cavity.
The total force induced by a global motion of the cavity may be evaluated in the quasistatic limit, {\it i.e.} for slow motions corresponding to frequencies lower than the cavity modes.
The motional force  exerted by vacuum fields on the cavity may then be written as an expansion
in the frequency 
\beqa
\label{global_motion_force}
&&<\delta F[\omega]> = \lbrace\mu \omega^2 + ... \rbrace \delta q[\omega]\nonumber\\ 
&&<\delta F(t)>= -\mu {d^2\over dt^2}\delta q(t) + ...\qquad \mu ={1\over2}\sum_{ij}\partial_\omega^2\chi_{F_i F_j}[0]
\eeqa
As expected, at zero frequency, the force (\ref{global_motion_force}) induced by a quasistatic motion of the mirrors corresponds to a variation of the mean Casimir force (\ref{Casimir_force}) due to the variation of the length of the cavity, so that it cancels in the total force. In conformity with Lorentz invariance of vacuum,  the mirrors velocities do not contribute so that next order contribution corresponds to the mirrors' accelerations.  
At lowest frequency order, the total motional force felt by the cavity (\ref{global_motion_force}) then depends on its global acceleration and may be seen as a correction to the mass of the cavity.
 This mass correction may be expressed 
in terms of the Casimir energy $E_C$ and the mean Casimir force $F_C$ \cite{JR93b}
\beqa
\label{mass_correction}
\mu =\frac{E_C- F_C q}{c^ 2 }
\eeqa
Dependence on $c$, the light velocity, has been restored for illustrative purposes. 
One can see that the mass correction (\ref{mass_correction}) precisely corresponds to the contribution of energy to inertia in the case of a stressed rigid body, as required by the law of inertia of energy according to relativity theory \cite{Einstein07}. 
In fact, the law of inertia  may be shown to express the 
conservation of the symmetry generator associated with Lorentz boosts \cite{Einstein06}.
The mass correction (\ref{mass_correction}) generalizes this law to include vacuum energies.

Although performed on a model of a cavity which cannot be considered as complete
(in particular, a complete description should be performed in terms of renormalized quantities), the previous discussion already allows one to conclude that field fluctuations modify the inertial response of a scatterer moving in vacuum. 
Moreover, the motional effects are compatible with the principles of relativity and may be consistently computed within the framework of linear response theory.
Due to its relation with the fluctuations of quantum fields,  the dissipative part of the motional response satisfies the same space-time symmetries as the vacuum state.
In fact, the reactive part of the motional response also conforms to these symmetries. 
As a result, the law of inertia of energy also applies to Casimir energies, which are 
a special case of a contribution of vacuum fields to energy (\ref{Casimir_force}).

\section{Mass as a quantum observable}

The study of the motion of a cavity embedded in quantum field fluctuations confirms the relativistic relation between mass and energy. It also shows that scattering of field fluctuations  modifies the mass of a scatterer. This property follows from the principles of relativity theory  relating motion with space-time symmetries, when applied within the linear response formalism (\ref{linear_response}, \ref{displacements}).
There result fundamental modifications of the status of mass.
Quantum fluctuations and symmetries impose to abandon the classical representation of mass as a mere characteristic parameter and to use the same representation as for all physical observables, {\it i.e.} as a quantum operator. 

\subsection{Quantum fluctuations of mass}

The  mass induced by vacuum field fluctuations (\ref{global_motion_force}, \ref{mass_correction}) may be seen as the energy  taken on field fluctuations due to time delays induced by scattering (\ref{phase_shift}, \ref{Casimir_force}).
The mass correction not only depends on time delays associated with the scattering matrix but also on
the energy density of incoming quantum field fluctuations. Following the same line of approach, motion in a thermal bath of quantum fields can be shown to induce, besides the well-kown friction force,  a mass correction which depends on the bath temperature \cite{JR93c}.

The dependence of a scatterer's mass on the energy density of quantum field fluctuations shows that mass cannot be considered any more as a mere parameter characterizing the scatterer.
The scatterer's mass inherits the quantum properties which are associated with the fluctuations of  scattered fields and hence must be reperesented by a quantum operator. The  contribution
induced by scattered fields leads to  quantum fluctuations of mass which can also be characterized by correlation functions (\ref{KMS_relations})
\beqa
\label{mass_fluctuations}
&&C_{MM}(t-t^\prime) \equiv<M(t) M(t^\prime)> - <M(t)> <M(t^\prime)>\nonumber\\
&&C_{MM}(t) \equiv \int_{-\infty}^\infty {d \omega \over2\pi} e^{-i\omega t} C_{MM}[\omega] 
\eeqa
 Correlation functions of the mass operator can be deduced from the linear response formalism
used to obtain motional responses in vacuum. Mass correlation functions  follow from the scattering matrix of quantum input fields (\ref{2d-fields}, \ref{scattering_matrix}) and from correlation functions of their energy-momentum tensor (\ref{stress_tensor}). 

In order to extend the analysis beyond the case of a cavity and to discuss simple  explicit expressions, we consider again the model of a scatterer as in previous sections (\ref{2d-fields}, \ref{scattering_matrix}), but now written under the form  of an explicitly relativistic  Lagrangian ${\cal L}$ \cite{JR93d}
\beqa
\label{Lagrangian}
&&{\cal L}= {1\over2}(\partial_t\phi^2 -\partial_x\phi^2) + \int ds \sqrt{1-\dot{q}(s)^2}\delta(x-q(s)) M, \qquad \dot{q}\equiv{d q\over ds}\nonumber\\
&&M \equiv M_0 + \Omega \phi^2(q)
\eeqa
 $\phi$ is a scalar field propagating in a 2-dimensional space-time and $q$ the space-time trajectory,
 parametrized by $s$, of a point-like scatterer. 
The scatterer's mass $M$ is the sum of a bare mass $M_0$
 and of an interaction term localized on the scatterer's trajectory. 
The bare coupling $\Omega$ can also be seen to play the role of a frequency cut-off.
Lagrangian (\ref{Lagrangian}) describes a general relativistic interaction between a scalar field and a pointlike particle,
with the further assumption of a quadratic interaction (the following arguments apply to general forms of interaction with minor
complications). 
In the limit of a scatterer at rest, the Lagrangian (\ref{Lagrangian}) being quadratic in the fields, the scattering matrix (\ref{scattering_matrix}) associated with the interaction is easily obtained, providing the corresponding  phase-shift or time delay 
\beqa
\label{time_delays}
&&s[\omega] = 1 + r[\omega], \qquad r[\omega] = -{\Omega\over\Omega -i\omega}\nonumber\\
&&2\tau[\omega] =\partial_\omega \Delta[\omega] =
{2\Omega \over \Omega^2 + \omega^2}
\eeqa
Energy conservation for the system (\ref{Lagrangian}) shows that the scatterer's mass undergoes a correction $\mu$ which is indeed related to the frequency dependent time delay, as for a cavity
\beqa
\label{mass_correction2}
&&\mu \equiv <\Omega\phi(q)^2> =  \int_0^\infty {d\omega \over 2\pi} 
\hbar \omega \tau[\omega]
\eeqa
The mass correction (\ref{mass_correction2}), when written in terms of time delays (\ref{time_delays}) is infinite, due to an ultraviolet divergence. A counterterm must be added to the bare coupling to compensate this divergence. In fact, expression (\ref{time_delays}) is a crude approximation of the scattering matrix associated with the system (\ref{Lagrangian}), which is only valid when the scatterer remains approximately at rest during scattering, {\it i.e.} for rather low field frequencies. For fields with an energy of the order of the scatterer's mass, recoil cannot be ignored and the scattering matrix, hence the time delay,  differs significantly from (\ref{time_delays}). In a self-consistent treatment of infinities one must come back to the general definition (\ref{evolution_operator},\ref{S-matrix}) of the scattering matrix and consider a renormalization of physical observables. But the  approximation (\ref{time_delays}) of the scattering matrix appears sufficient to exhibit the essential quantum properties of the mass observable.

The induced mass (\ref{mass_correction2}) depends on the energy-density of vacuum fields, $\mu$ corresponding to its mean value, so that it also exhibits quantum fluctuations.
Using the simple model (\ref{Lagrangian}, \ref{time_delays}), one derives the corresponding quantum fluctuations in the frequency domain (\ref{mass_fluctuations}), which can also be written in terms of time delays 
\beqa
\label{mass_quantum_fluctuations}
C_{MM}[\omega] = 2\hbar^2 \theta(\omega)
\int_0^\omega {d\omega^\prime \over 2\pi}
\omega^\prime(\omega-\omega^\prime)\tau[\omega^\prime] \tau[\omega-\omega^\prime]
\eeqa
In spite of approximations which have been made, expression (\ref{mass_quantum_fluctuations})  for 
the correlations of mass quantum fluctuations remains valid for  
frequencies which are smaller than the frequency cut-off $\Omega$.
One recovers for the mass observable the characteristic spectrum of quantum fluctuations in vacuum: 
the factor proportional to  $\theta(\omega)$  reduces excitations to positive frequencies only. This results, in the time domain, into a non vanishing commutator for the mass observable at different times, 
implying that mass must be represented by a quantum operator. 
The quantum properties of the mass observable manifest themselves at short time scales.
Indeed, extrapolating (with the restrictions which have been formulated) this simple model to high frequencies, one observes that mass fluctuations cannot be neglected due contributions at short times 
\beqa
<M^2> - <M>^2  =  2 <M>^2
\eeqa
On another hand, the fluctuations of the mass observable are negligible in the low
frequency domain. For the simple model (\ref{Lagrangian}), variations behave as the third power of frequency
\beqa 
C_{MM}[\omega] \sim {\hbar^2 \over 6 \pi} \theta(\omega)
{\omega^3\over\Omega^2} \qquad \qquad {\rm for} \qquad \omega \ll \Omega
\eeqa
This last  property of the mass observable justifies its approximation by a constant parameter,
as long as low frequency motions are considered. But fluctuations which come into play at higher frequencies limit the validity of this approximation \cite{JR93d}.
This means that in a complete treatment, the remormalized mass of a scatterer is obtained under the form of a quantum operator. In other words, the renormalized mass of a scatterer follows from the renormalized energy-momentum tensor of scattered fields. The value of the parameter defining the renormalization prescription corresponds to the mean value of the renormalized mass observable. 

\subsection{Mass and conformal symmetries}

The linear response formalism confirms, within the quantum framework, the relativistic relation between motion and space-time symmetries.
Due to this relation, mass cannot be considered as an ordinary quantum observable. The quantum operator associated with mass must remain consistent with the general identification of generators of space-time symmetries with constants of motion. 
The relativistic relation between energy and inertial mass induced by acceleration has been seen in previous section to hold in a quantum framework and to include the energy due to vacuum fluctuations. This relation should also possess a general expression in terms of quantum operators associated with space-time symmetries.
It appears that this property is ensured by the existence of a group of symmetries which extends the Poincar\'e group of translations and Lorentz transformations to include symmetries associated with accelerations.

In the following, we discuss the specific properties relating mass and acceleration in a quantum framework,
only considering the case of a flat four-dimensional space-time.
In a relativistic quantum theory, the generators of spacetime symmetries describe changes of reference frame and also correspond to quantities which are preserved by the equations of motion. 
The transformations of quantum observables under translations and Lorentz transformations are respectively given by their commutators with the energy-momentum $(P_\mu)_{\mu=0,1,2,3}$ and the angular momentum $(J_{\mu \nu })_{\mu,\nu=0,1,2,3}$.
The actions of these generators on observables  satisfy the Poincar\'e algebra, {\it i.e.} the following commutation relations
\beqa
\label{Poincare}
&&[P_\mu ,P_\nu] = 0\nonumber\\
&&[ J_{\mu \nu }, P_\rho ] = i\hbar\left( \eta _{\nu \rho }P_\mu 
 -\eta _{\mu \rho }P_\nu\right)\nonumber\\
&&[J_{\mu \nu }, J_{\rho \sigma }] = i\hbar\left(
 \eta _{\nu \rho } J_{\mu \sigma } + \eta _{\mu \sigma }J_{\nu \rho } 
-\eta _{\mu \rho } J_{\nu \sigma} - \eta _{\nu \sigma }J_{\mu \rho }\right)  
\eeqa
$\eta_{\mu\nu} \equiv {\rm{diag}}(1,-1,-1,-1)$ denotes Minkowski metric tensor and determines the light cones  defining the causal structure of space-time.
Propagations of electromagnetic fields and of gravitation fields, at the linearized level, follow these light cones so that the corresponding solutions of the equations of motion form a representation of the group of light cone symmetries \cite{Bateman09,Cunningham09}.
These correspond to conformal symmetries which include, besides the generators of Poincar\'e algebra (\ref{Poincare}),
generators corresponding to a dilatation $D$ and to transformations to accelerated frames $(C_\mu)_{\mu,\nu=0,1,2,3}$.
The corresponding generators satisfy commutation rules which define the conformal algebra
\beqa
\label{conformal_algebra}
&&[D, P_\mu] = i\hbar P_\mu   \qquad \qquad 
[D, J_{\mu \nu }] = 0\nonumber\\
&&[P_\mu , C_\nu] = -2i\hbar \left(\eta _{\mu \nu }D-J_{\mu \nu }\right), \qquad 
[J_{\mu \nu }, C_\rho] = i\hbar\left(\eta _{\nu \rho }C_\mu 
 -\eta _{\mu \rho }C_\nu\right)\nonumber\\
&&[D, C_\mu] = - i\hbar C_\mu\nonumber\\
&&[C_\mu , C_\nu] = 0
\eeqa
The generators of the conformal algebra (\ref{conformal_algebra}) describe space-time symmetries associated with light cones and field propagation \cite{Bracken71,Bracken81}. Hence, they  correspond to symmetries of the vacuum state and of motional responses in vacuum fields. They are satisfied in particular by the radiation reaction force in vacuum, as discussed in previous section. 
As shown in the following, they also allow to analyse the relation between conformal generators and motion in terms of quantum observables. For that purpose, 
 one must first define the quantum observables which can be associated with positions in space-time. 

According to the relativistic conception, positions in space and time should be defined as physical observables \cite{Einstein05b}.
Time is delivered by special systems designed for that purpose, {\it i.e.} clocks. 
A given set of synchronized clocks and emitters, disseminating time references along propagating signals 
(using electromagnetic fields for instance), builds a reference system which allows one to determine coordinates in space and time. 
The several time references  received at a given location determine the positions of this location with respect to the reference system. 
This notion of space-time, based on a realization of positions  by means of observables, 
is implemented nowadays in metrology  \cite{Guinot77} and in reference systems used for positioning 
and for navigation around the Earth and in the solar system \cite{IEEE91,Bahder01}.
However, this  implementation differs from the representations of space and time which are used in quantum field theories. 
In order to maintain the same status for space and time, positions are similarly represented in quantum field theories as mere parameters.
These parameters describe a classical manifold on which quantum fields and physical observables can be defined.
In such a representation, positions in space-time have lost the nature of observables as required by relativity theory.
But, nothing in principle prevents one from applying the relativistic definition of observables
describing space-time positions within the context of quantum field theory. 
The main tradeoff lies in the increased complexity of the observables representing positions in space-time,
with the significant advantage of restoring a consistency between the principles of relativity theory and the 
formalism underlying quantum theory. 

Following the relativistic approach, time references may be defined as observables built on the energy-momentum tensor 
of exchanged fields \cite{JR96a}.  Then, localization in space-time by means of quantum fields leads to the definition of
space-time positions as quantum observables $(X_\mu)_{\mu=0,1,2,3}$, built from quantities which are conserved by field propagation, {\it i.e.} from the generators of space-time symmetries \cite{JR96b}($\cdot$ denotes the symmetrized product of operators)
\beqa
\label{quantum_positions}
X_\mu = {1\over P^2}\cdot\left(P^\lambda \cdot J_{\lambda \mu} + P_\mu \cdot D\right)
\eeqa
Space-time positions are represented by operators (\ref{quantum_positions}) which belong to an extension of the enveloping algebra
of the Lie algebra of conformal symmetries (\ref{conformal_algebra}). In particular, the definition of space-time positions excludes massless field configurations ($P^2\equiv M^2 =0$), {\it i.e.} its realization requires configurations involving fields propagating in different directions. Then, positions belong to an algebra which is generated by quantum fields and which contains quantum observables and relativistic frame transformations. 
Positions as defined by (\ref{quantum_positions}) are also conjugate to  energy-momentum observables
\beqa
\label{canonical_commutator}
 [P_\mu , X_\nu] = - i\hbar \eta_{\mu \nu}
\eeqa
One must note at this point that equations (\ref{quantum_positions}) define quantum operators which describe positions not only in space but also in time. Furthermore, the time operator thus defined is conjugate to the energy observable according to (\ref{canonical_commutator}). As these operators are built on quantum fields which possess a state with minimal energy, the vacuum,
all conditions seem satisfied to apply a well-known theorem \cite{Pauli33}, in its relativistic formulation \cite{Wightman62}, stating the impossibility to define such an operator.
In fact, one can see that a further condition assumed by the theorem, namely the self-adjointness of the time operator, is not fulfilled here.
Although this property is often satisfied by quantum observables, it does not appear to be necessary in general \cite{Bogolubov75}. Observables with real eigenvalues only require to be represented by hermitian operators and  the definition domains of an operator and its adjoint
may differ. This happens in particular when part of the Hilbert space must be excluded from the definition domain, as is the case for the time operator defined by (\ref{quantum_positions}). 
The exclusion of massless field configurations then allows one to escape the objection raised by 
the theorem and to define a time operator satisfying the required commuation rules with the generators of space-time symmetries \cite{JR96a}. 

It can be seen that positions thus defined (\ref{quantum_positions}) tranform according to classical rules under rotations and dilatation
\beqa
{i\over\hbar}[J_{\mu \nu} , X_\rho] = \eta_{\mu \rho} X_\nu 
 - \eta_{\nu \rho} X_\mu, \qquad  {i\over\hbar}[D , X_\mu ] = X_\mu
\eeqa
Transformations to uniformly accelerated frames are then given by the 
conformal generators $C_\mu$ (\ref{conformal_algebra}). 

In conformity with relativity theory \cite{Einstein05b}, the mass observable $M$ is  a Lorentz invariant (Poincar\'e invariant) built on energy-momentum $P_\mu$ 
\beqa
\label{mass}
M^2 = P^\mu P_\mu, \qquad {i\over\hbar}[P_\mu , M^2] = {i\over\hbar}[J_{\mu \nu }, M^2 ] = 0
\eeqa
But the extension of space-time transformations to the conformal algebra, and in particular the action of the dilatation operator,
 shows that the mass observable cannot be considered as a mere parameter.
Mass (\ref{mass}) and the conformal generators (\ref{conformal_algebra}) are  embedded 
within the same algebra of observables, with mass loosing its invariance property
\beqa
\label{mass_dilatation}
{i\over\hbar}[D , M] = -M
\eeqa

The  mass operator (\ref{mass},\ref{mass_dilatation})  can also be seen to provide an extension to the quantum framework of the  relativistic relation between positions (\ref{quantum_positions}) and the law of inertial motion. 
The transformation of mass (\ref{mass}) under a uniform acceleration is obtained from the conformal algebra (\ref{conformal_algebra}) and is seen to involve the position observable
\beqa
\label{accelerated_mass}
 {i\over\hbar}[C_\mu , M ] = -2 M \cdot X_\mu
\eeqa
The transformation (\ref{accelerated_mass}) of the mass observable takes the same form as the classical
red-shift law describing the effect of acceleration on frequencies \cite{Einstein07}.
Rewriting equation (\ref{accelerated_mass}) as the action of the generator $\Delta$ corresponding to an  acceleration $a^\mu$
on the mass observable $M$, the result identifies with Einstein law, now written in terms of quantum positions $X_\mu$
\beqa
\label{red-shift}
\Delta \equiv \frac{a^\mu}2 C_\mu, \qquad {i\over\hbar}[\Delta , M ] = - M \cdot \Phi, \qquad \Phi = a^\mu X_\mu
\eeqa
The transformation (\ref{red-shift}) of the mass is proportional to the mass itself and to a potential $\Phi$ which is given by the product of the acceleration with the quantum position. This quantum version of the classical
red-shift law describes the effect on frequencies of an acceleration or an equivalent gravitational potential $\Phi$,
 according to relativity theory \cite{Einstein07}.
Conversely, the transformation of mass under accelerations (\ref{accelerated_mass}), given by the conformal algebra,
can be used to define quantum positions, expressions (\ref{quantum_positions}) being then recovered \cite{JR97a}.

The representation of frame transformations and motions as actions of generators of symmetries ensures that transformations of observables remain compatible both with relativity and quantum theory. Expression (\ref{red-shift}) makes it explicit
in the case of the transformation of mass under a uniform acceleration, which appears to be 
 equivalent to a uniform gravitational potential.
The consistency of both interpretations is ensured in this case 
by the relation made in the quantum framework between frequency    
and energy, more precisely, by the conformal invariance of Planck constant $\hbar$.
This property holds more generally, the particular form (\ref{red-shift}) of the gravitational potential in terms of quantum positions appearing as a particular case of a quantum generalization of the metric field. Using conformal symmetry, the covariance rules which, in classical theory, implement the equivalence between motion (or changes of reference frames) and gravitation (or metric fields)
may be given a generalized form which applies to quantum observables \cite{JR97a}.
Symmetries and their associated algebras then provide a way to extend to the quantum framework the equivalence principle which lies at the heart of general relativity \cite{Einstein07}.

\section{Metric extensions of GR}

The previous part has shown that quantum fluctuations modify the relation of mass to
inertial motion and gravitation.  
In this part, we discuss a similar  modification of gravitation which is due to  quantum fluctuations
of stress tensors and leads to effects which might be observable at the macroscopic level.

In order to  discuss this issue, one must first come back to the founding principles of GR. 
First, the equivalence principle, in its weak version, states the universality of free fall and gives GR its geometric nature.
Violations of the equivalence principle are constrained by modern experiments to remain extremally small, below the $10^{-12}$ level, so that the equivalence principle is one of the best tested properties of nature \cite{Will06}.
The level of precision attained by tests, at least for scales ranging from the submillimiter to a few A.U., disfavors strong violations of the equivalence principle, so that modifications of GR should first be looked for among theories which still obey this principle. 

Then,  GR may be characterized, as a field theory, by the coupling it assumes  between
gravitation (or metric fields) and sources. This coupling is equivalent to the gravitation equations which determine the metric
tensor from the distribution of energy-momentum in space-time.
According to GR, the curvature
tensor of the metric and the energy-momentum tensor of sources are in a simple relation \cite{Einstein15,Hilbert,Einstein16}.
Einstein curvature tensor $E_{\mu\nu}$ is simply proportional to the energy-momentum tensor 
$T_{\mu\nu}$ and a single constant, Newton gravitation 
constant $G_N$, describes  gravitational coupling
\beqa
\label{Einstein_Hilbert}
E_{\mu\nu} \equiv R_{\mu\nu}-{1\over2}g_{\mu\nu} R={8\pi G_N \over c^4}T_{\mu\nu}
\eeqa
$R_{\mu\nu}$ and $R$ denote the Ricci and scalar curvatures which are built on Riemann curvature tensor. 
 As a result of Bianchi identities, Einstein curvature tensor  has a null covariant divergence, like the energy-momentum tensor,
so that equation (\ref{Einstein_Hilbert}) makes a simple connection between a geometric property of curvatures and the physical law of energy-momentum conservation. The latter property identifies with the geodesic motion describing free fall. 
But these properties could still be satisfied by fixing other relations between  
curvature and energy-momentum tensors, so that Einstein-Hilbert choice is the simplest but not the only physical possibility.

Indeed, as discussed in the following, even if gravitation is described by GR at the classical level,
quantum fluctuations of stress tensors lead to modifications of gravitation equations (\ref{Einstein_Hilbert}),
while preserving the metric nature of the theory.  
 
\subsection{Radiative corrections}

Form now on, we focus on gravitation theories which preserve  the equivalence principle, {\it i.e.} metric theories.
Space-time is then represented by a four-dimensional manifold, endowed with a metric $(g_{\mu\nu})_{\mu,\nu=0,1,2,3}$,
with Minkowskian signature, identifying with gravitation fields. Gravitation may be treated as a field theory which is characterized by its Lagrangian, giving equations of motions for the gravitation fields such as Einstein-Hilbert equations  (\ref{Einstein_Hilbert}) for GR
\cite{Weinberg72}.

Einstein-Hilbert equations (\ref{Einstein_Hilbert}) describe the propagation of gravitation fields in empty space
in the classical framework of GR. But, if treated on the same footing as fields corresponding to other fundamental interactions,
gravitation fields must also possess quantum fluctuations which are induced by quantum fluctuations of sources, {\it i.e.} of energy-momentum tensors \cite{Thirring61,Feynman63,Weinberg65}.
 Quantum fluctuations lead to effective equations for the propagation of gravitation fields which are modified,
with consequences which may remain significant in the classical limit.
To make this more explicit, it is convenient to first consider gravitational fluctuations in flat space.
Fluctuations of gravitation fields are then represented as perturbations ($(h_{\mu\nu})_{\mu,\nu=0,1,2,3}$) of Minkowski metric, which may equivalently be 
 written as functions of position 
in spacetime or of a wavevector in Fourier space
\beqa
&&g_{\mu\nu} = \eta_{\mu\nu} + h_{\mu\nu}, \qquad\eta_{\mu\nu} = {\rm diag}(1, -1, -1, -1) 
\quad,\quad |h_{\mu\nu}| << 1\nonumber\\
&&h_{\mu\nu}(x) \equiv \int {d^4 k \over (2\pi)^4}e^{-ikx} h_{\mu\nu}[k]
\eeqa
The definition of metric fields suffers from ambiguities related to the choice of coordinates
but, at the linearized level, gauge invariant fields are provided by Riemann, Ricci, scalar and Einstein curvatures  
\beqa
&&R_{\lambda\mu\nu\rho}[k] = {1\over2}\lbrace
k_\lambda k_\nu h_{\mu\rho}[k] - k_\lambda k_\rho h_{\mu\nu}[k]
- k_\mu k_\nu h_{\lambda\rho}[k] + k_\mu k_\rho h_{\lambda\nu}[k]\rbrace\nonumber\\
&&R_{\mu\nu} = {R^\lambda}_{\mu\lambda\nu}\quad, \quad R = {R^\mu}_\mu\quad,\quad
E_{\mu\nu} = R_{\mu\nu} -\eta_{\mu\nu}{R\over2}
\eeqa
Classically, metric fields are determined from energy-momentum sources by Einstein-Hilbert equations  
of GR (\ref{Einstein_Hilbert}) which, at the linearized level and in the momentum domain, take a simple form
\beqa
\label{linearized_EH}
E_{\mu\nu}[k] = {8\pi G_N\over c^4} T_{\mu\nu}[k]
\eeqa
Equations of motion (\ref{linearized_EH}) in fact describe the coupling between metric fields and the total energy-momentum tensor of all fields, {\it i.e.} the corresponding coupling terms in their common Lagrangian. Due to the non linear character of  gravitation theory,  these equations  include the energy-momentum tensor of gravitation itself \cite{Weinberg72}. Equations (\ref{linearized_EH}) determine the metric fields which are generated by classical gravitation sources and may be seen as describing the response of metric fields to  energy-momentum tensors, when quantum fluctuations are ignored. However, virtual processes associated with quantum fluctuations, {\it i.e.} radiative corrections,  must be taken into account  when solving equations (\ref{linearized_EH}). There result modifications of  the graviton propagator or of the effective coupling between gravitation and its sources  \cite{deWitt62,Deser74,Capper74}.
It is well-known that radiative corrections associated with Einstein-Hilbert equations (\ref{linearized_EH}) involve divergences which cannot be treated by usual means, due to the non renormalizability of GR \cite{deWitt62,tHooft74} . However, these corrections result in embedding GR within a larger family of gravitation theories, with Lagrangians involving not only the scalar curvature but also  quadratic forms in Riemann  curvature. These theories appear to be renormalizable and to constitute reasonable  extensions of GR.
Furthermore, this enlarged family shows particular properties with respect to renormalization group trajectories \cite{Reuter02,Fradkin82,Hamber07}, which hint at a consistent definition of a gravitation theory, with GR being a very good approximation within the range of length scales where it is effectively observed. 
Hence, we shall consider GR as an approximate effective theory and shall focus on the corrections to GR which remain 
to be taken into account in the range of length scales where GR is very close to the actual gravitation theory. 
It is also usually objected that gravitation theories with equations of motion involving higher derivatives of metric fields
lead to violations of unitarity, or unstability problems, which are revealed by the presence of ghosts.
We shall note that arguments have been advanced for denying to consider these objections as real dead ends \cite{Simon90,Hawking02}.
Here, we shall just take the minimal position of restricting attention to a range of scales where both the gravitation theory  remains close to GR and  instabilities do not occur.

Keeping in mind the previous restrictions, one may see the modified gravitation propagator, including the effect of radiative corrections, as an effective response function of metric fields to energy-momentum tensors \cite{JR95b}.
Gravitation equations then take the generalized form of a linear response relation between Einstein curvature and  energy-momentum tensors
 \beqa
\label{gravitation_law}
&&E_{\mu\nu} [k] = \chi_{\mu\nu}^{\lambda\rho} [k] ~ T_{\lambda\rho} [k]  = \left( {8\pi G_N\over c^4}\delta^\lambda_\mu\delta^\rho_\nu + \delta \chi_{\mu\nu}^{\lambda\rho} [k] \right) T_{\lambda\rho} [k]
\eeqa
The effective gravitation equations (\ref{gravitation_law}) take the same form as response functions induced by motion
(\ref{linear_response},\ref{displacements}), with energy-momentum tensors playing the role of displacement generators.
This property follows from the relation made in relativity theory between changes of coordinates and motion 
(this relation in particular provides the definition of a symmetric energy-momentum tensor \cite{Landau84}). 
Assuming that corrections induced by quantum fluctuations correspond to  perturbations,
the effective gravitation equations  (\ref{gravitation_law}) should appear as a perturbation of Einstein-Hilbert equations (\ref{Einstein_Hilbert}). These perturbations  may be captured in a function $\delta\chi_{\mu\nu}^{\lambda\rho}$,
which may be seen as a momentum dependent correction to the coupling constant $G_N$, or as a  nonlocal correction to the gravitational coupling in the space-time domain. 

Studying the coupling between gravitation and  different fields \cite{Capper74,JR95b} allows one to
derive some general properties of the modification brought by radiative corrections to Einstein-Hilbert equations,
even if a complete determination of the modified response function $ \chi_{\mu\nu}^{\lambda\rho}$ is not available. 
The different quantum fields coupling to gravitation lead to radiative corrections which differ in  two sectors with
different conformal weights.
Conformally invariant fields, such as electromagnetic fields, which correspond to traceless energy-momentum tensors 
only affect the conformally invariant sector, corresponding to Weyl curvatures.
On the other hand, massive fields, which involve energy-momentum tensors with a non vanishing trace, modify both sectors. 
The two sectors then correspond to different modifications of the gravitational coupling constant $G_N$
into running coupling constants, so that $G_N$ should be replaced by two slightly different running coupling constants.
In the linearized approximation, and in the case of a static pointlike source, the equations for metric fields  (\ref{gravitation_law}) 
may be rewritten in terms of projectors on transverse components
\beqa
\label{linearized_GE}
&&T_{\mu\nu}= \delta_{\mu 0} \delta_{\nu 0} T_{00}, \qquad T_{00}[k] =Mc^2\delta(k_0)\nonumber\\
&&E_{\mu\nu}= E^{(0)}_{\mu\nu} + E^{(1)}_{\mu\nu}, \qquad \pi_{\mu\nu}[k]\equiv \eta_{\mu\nu} -{k_\mu k_\nu\over k^2}\nonumber\\
&&E^{(0)}_{\mu\nu} = 
\left( \pi _{\mu}^{0}\pi_{\nu}^{0}-{\pi_{\mu\nu}\pi^{00}\over3}\right)
\, \frac{8\pi G^{(0)}}{c^{4}}T_{00}, \qquad 
E^{(1)}_{\mu\nu} =  \frac{\pi _{\mu\nu}\pi ^{00}}{3}
\frac{8\pi G^{(1)}}{c^{4}}T_{00} \nonumber\\
&&G^{(0)}[k]= G_N  + \delta G^{(0)}[k], \qquad
G^{(1)}[k]= G_N  + \delta G^{(1)}[k]
\eeqa
At the linearized level, the decomposition on the two sectors with different conformal weights ($E^{(0)}$ and $E^{(1)}$)
is easily performed in the momentum domain by means of projectors.
The modified equations for gravitation then take the same form as Einstein-Hilbert equations (\ref{linearized_EH}), but in terms of two running coupling constants $G_N+\delta G^{(0)}$
and  $G_N+\delta G^{(1)}$, which depend on the momentum or length scale and which slightly modify Newton gravitation constant $G_N$ \cite{JR05a,JR05b}.
These two scale dependent couplings describe gravitation theories which remain close to GR within a certain range of scales along renormalization trajectories. They thus constitute a neighborhood of GR
made of a large collection of theories labelled by two functions of an arbitrary scale parameter.  
As discussed above, GR just appears as particular point in this neighborhood which is compatible with the observations made on our gravitational environment in an  accessible range of scales.

\subsection{Anomalous curvatures}

For the sake of simplicity, modifications of Einstein-Hilbert equations induced by radiative corrections have been 
presented, in  previous section, within the context of a linearized gravitation theory around a flat space-time (namely for weak gravitation fields and at first order in such fields).
But the same mechanisms are easily seen to occur in the case of any space-time, endowed with an arbitrary background metric field, leading to modifications of Einstein-Hilbert equations  (\ref{Einstein_Hilbert})  which may still be described as in equations (\ref{gravitation_law}) by a general response of metric fields to energy-momentum tensors.
Again,  arguments derived from observations of our gravitational environment entail that gravitation equations should remain close to Einstein-Hilbert equations and take the following form
\beqa
\label{modified_EH}
&&E_{\mu\nu}  = \chi_{\mu\nu}^{\lambda\rho} \star T_{\lambda\rho}   = {8\pi G_N\over c^4}T_{\mu\nu}  + \delta \chi_{\mu\nu}^{\lambda\rho}\star T_{\lambda\rho} 
\eeqa
The $\star$ product denotes a convolution in space-time which replaces the ordinary product in the momentum domain. The effective response then introduces a non local relation between Einstein curvature (with its full non linear dependence on metric fields) and energy-momentum tensors. The perturbed response function still differs in two sectors with different conformal weights and is thus equivalent to two different running coupling constants $G_N+\delta G^{(0)}$
and  $G_N+\delta G^{(1)}$. Thus, also at the full non linear level, GR  appears as embedded within a large family of gravitation theories labelled by two functions of a length scale parameter. Although the two running coupling constants remain close to Newton gravitation constant $G_N$,
the relation they induce in general  between curvatures  and energy-momentum tensors is not explicit, due to an interplay between non linearity and non locality \cite{JR06a}. However, for discussing the observable consequences entailed by the modified gravitation equations (\ref{modified_EH}), one only needs the solutions of these  equations corresponding to given energy-momentum distributions. In that case, the modified equations (\ref{modified_EH}) remaining close to Einstein-Hilbert equations   (\ref{Einstein_Hilbert}), their solutions are  small pertubations of the metric solutions satisfying Einstein-Hilbert equations.

The metric solution of generalized gravitation equations (\ref{modified_EH}) lies in the vicinity of GR metric and satisfies perturbed equations
 (in the following, the notation  $\stand{\quad}$ stands for a GR solution of Einstein-Hilbert equations (\ref{Einstein_Hilbert}))
\beqa
\label{anomalous_Ricci}
&&E_{\mu\nu}(x) \equiv \stand{E_{\mu\nu}}(x) + \delta E_{\mu\nu}(x) \,,\qquad  \stand{E_{\mu\nu}}(x) = 0 \quad {\rm{when}} \quad T_{\mu\nu}(x) = 0\nonumber\\
&&\delta E_{\mu\nu}(x) \equiv \int d^4 x'\, \delta \chi^{\lambda\rho}_{\mu\nu}(x,x') 
T_{\lambda\rho}(x')\nonumber\\
&&\delta E_{\mu\nu} = \delta E^{(0)}_{\mu\nu} + \delta E^{(1)}_{\mu\nu}
\eeqa
Solutions of generalized equations (\ref{modified_EH}) then identify with anomalous Einstein or Ricci  curvatures, {\it i.e.} with metrics leading to Ricci components which do not vanish in empty space (outside gravitational sources), contrarily to GR solutions.
When considering solutions to the gravitation equations of motion (\ref{modified_EH}), with the latter corresponding to quantum perturbations of Einstein-Hilbert equations, the two running coupling constants $\delta G^{(0)}$ and $\delta G^{(1)}$ which modify Newton gravitation constant $G_N$  are seen to be equivalent to anomalous Einstein curvatures (\ref{anomalous_Ricci}). The latter possess two independent components only, which may be chosen as the  components of Einstein curvature with conformal weight $0$,
 $\delta E^{(0)}$ related to Weyl curvature, and with conformal weight $1$, $\delta E^{(1)}$ equivalent to the scalar curvature.
The relation between coupling constants and anomalous curvatures is obtained by extending the linearized limit (\ref{linearized_GE}) \cite{JR06a}.

For applications which will be our main concern here, namely gravitation in the outer part of the solar system, it will be suffcient to consider the  stationary and isotropic case. Using Schwartzschild coordinates, a stationary and isotropic metric $g_{\mu\nu}$ may be written (with $t$, $r$ and $\theta$, $\varphi$ denoting the time, radial and angular coordinates respectively)
\beqa
\label{isotropic_metric}
g_{\mu\nu} dx^\mu dx^\nu \equiv g_{00}c^2dt^2 - g_{rr} dr^2 -r^2(d\theta^2 + {\rm{sin}}^2\theta d\varphi^2)
\eeqa 
A stationary isotropic metric is then characterized by two functions of the radial coordinate $r$ only, namely its temporal and radial components
$ g_{00}$ and $ g_{rr}$ respectively. For the sake of simplicity, we shall also ignore effects due to the size and rotation of the gravitational source (these can be introduced without qualitatively changing the following discussions) and consider a
pointlike gravitational source. The corresponding GR solution may be written in terms of a single Newtonian potential $\Phi_N$
\beqa
\label{GR_metric}
&&\stand{E_\mu^\nu}(x) =  8\pi \kappa \delta^0_\mu \delta_0^\nu \delta^{(3)}(x), \quad \kappa \equiv {G_N M \over c^2}\nonumber\\
&&\stand{g_{00}}(r) = 1+ 2\Phi_N=  -{1\over\stand{g_{rr}}(r)},  \qquad \Phi_N\equiv -\kappa u, \qquad u\equiv{1\over r}
\eeqa
The metric given by (\ref{isotropic_metric}) and solution of the generalized gravitation equations (\ref{modified_EH}) lies in the vicinity of GR metric (\ref{GR_metric})
and corresponds to two anomalous Einstein curvature components (\ref{anomalous_Ricci}).
 In the stationary isotropic case, Ricci and Einstein curvatures have only two independent components, 
so that the anomalous components in the two different sectors, $\delta E^{(0)}$ and  $\delta E^{(1)}$,  may  be replaced by the 
temporal and radial components of Einstein curvature $\delta E^0_0$ and $\delta E^r_r$ (which may be rewritten in terms of Weyl and scalar curvatures). Then, solving for the metric field, 
the two  anomalous curvature components become equivalent to anomalous parts in the two 
components of the isotropic metric
\beqa
\label{anomalous_metric}
&&g_{00} = \stand{g_{00}} + \delta g_{00}, \qquad {\delta g_{00}\over\stand{g_{00}}} =\int {d u\over\stand{g_{00}}^2}
\int^u{\delta E_0^0\over u^4} d u^\prime 
+ \int{\delta E^r_r\over u^3}{d u\over \stand{g_{00}}}\nonumber\\
&& g_{rr} = \stand{g_{rr}}
+ \delta g_{rr}, \qquad {\delta g_{rr}\over\stand{g_{rr}}}= -{u\over\stand{g_{00}}}
\int{\delta E_0^0\over u^4} d u 
\eeqa
In case of a stationary point-like source, the two running coupling constants characterizing the modified gravitation equations (\ref{modified_EH})
are equivalent to two anomalous components of Ricci or Einstein curvature tensor (\ref{anomalous_Ricci}), or else to
two anomalous parts in the corresponding stationary isotropic metric (\ref{anomalous_metric}). If the first representation
better suits the framework of QFT, the last representation appears more appropriate to a study of the observable consequences 
of modified gravitation. 

For a phenomenological analysis, it is even more convenient to rewrite the two independent degrees of freedom under the form of two gravitation potentials. In the stationary isotropic case, the two sectors of anomalous components (\ref{anomalous_Ricci}) can be represented by two anomalous gravitational potentials, corresponding to  the temporal and radial components of Einstein curvature 
\beqa
&&\delta E^0_0 \equiv 2 u^4(\delta\Phi_N -\delta\Phi_P)^{\prime\prime}, \qquad   ()^\prime\equiv \partial_u\nonumber\\
&&\delta E^r_r \equiv 2 u^3\delta\Phi_P^\prime 
\eeqa	
Solutions for the metric components then provide the perturbation (\ref{anomalous_metric}) around GR solution (\ref{GR_metric})
in terms of anomalous potentials
\beqa
\label{extended_metric}
&&\delta g_{rr} = {2u\over(1+2\Phi_N)^2}(\delta\Phi_N-\delta\Phi_P)^{\prime} \nonumber\\
&&\delta g_{00} = 2\delta\Phi_N +4\kappa(1+2\Phi_N)
\int{u(\delta\Phi_N -\delta\Phi_P)^\prime - \delta\Phi_N\over(1+2\Phi_N)^2}d u
\eeqa
Equations (\ref{extended_metric}) in terms of two gravitational potentials $\Phi_N +\delta \Phi_N$ and $\delta \Phi_P$ provide metric extensions which remain close to GR
while  accounting for non linearities in the metric. They correspond to a modification $\delta \Phi_N$ of Newton potential $\Phi_N$
and to the introduction of a second potential $\delta \Phi_P$. The two gravitational potentials describe, up to combinations,  the two sectors introduced by the modified gravitation equations (\ref{modified_EH}) and span the corresponding family of gravitation theories labelled by two functions of a length scale,  which are equivalent to the two running coupling constants induced by radiative corrections \cite{JR06a}.

\subsection{Phenomenology in the solar system}

The family of extended metrics (\ref{GR_metric},\ref{extended_metric}), obtained by solving the generalized gravitation equations
(\ref{modified_EH}), provides a basis for a phenomenological analysis of gravitation in the neighborhood of a stationary pointlike source. Extended metrics (\ref{GR_metric},\ref{extended_metric}) depend on two functions (of a single variable, the distance to the gravitational source) which parametrize a vicinity of GR.
Observations performed in the gravitational field of the source should then allow one to characterize these two functions and thus to determine the nature of the theory describing gravitation in the neighborhood of a pointlike source.

In the solar system, tests of gravity are usually performed by comparing observations 
 with the predictions obtained from a family of  parametrized post-Newtonian (PPN) metrics which extend GR metric 
by introducing additional parameters \cite{Eddington57,WillNordtvedt72}.
In the approximation of a stationary pointlike gravitational source, ignoring effects due to size and  rotation,
PPN metrics may be  written in terms of a single potential $\phi$ (which takes in isotropic coordinates the same form as Newton potential $\Phi_N$ in Schwartzschild coordinates (\ref{GR_metric})) and on two constant parameters $\beta$ and $\gamma$ \cite{Will01}
\beqa
\label{PPN_metric}
&&g_{\mu\nu} dx^\mu dx^\nu \equiv g_{00}c^2dt^2 - g_{rr} \lbrace dr^2 +r^2(d\theta^2 + {\rm{sin}}^2\theta d\varphi^2)\rbrace\nonumber\\
&&g_{00} = 1 + 2 \phi + 2 \beta \phi^2 + \ldots, \qquad \phi = -{G_N M\over c^2 r}\nonumber\\
&&g_{rr} = -1 + 2 \gamma \phi + \ldots
\eeqa
Eddington parameters $\gamma$ and $\beta$ respectively
describe  linear effects on light deflection and non-linear effects on perihelia of planets.
They are defined so that GR corresponds to the particular values $\beta=\gamma=1$.
 
It is easily seen that PPN metrics (\ref{PPN_metric}) correspond to particular cases of the general metric extensions of GR which have been previously introduced
\beqa
\label{generalized_PPN_metric}
&&\delta\Phi_N = (\beta-1)\phi^2 + O(\phi^3), \qquad 
\delta\Phi_P = -(\gamma-1)\phi + O(\phi^2)\nonumber\\
&&\delta E^0_0 = {1\over r^2} O( \phi^2),\qquad
\delta E^r_r = {1\over r^2} \left(2 (\gamma-1) \phi +O( \phi^2) \right)\qquad \mathrm{[PPN]}
\eeqa
PPN metrics span a  two-dimensional subspace, labelled by $(\beta,\gamma)$,  of the neighborhood of GR corresponding to solutions of generalized gravitation equations (\ref{modified_EH}), which are labelled by two functions $\delta \Phi_N$ and $\delta \Phi_P$.
According to equations (\ref{generalized_PPN_metric}), this subspace corresponds to metrics with Einstein curvatures which vanish at large distances of the gravitational source. Alternatively, the  metric extensions of GR   (\ref{GR_metric},\ref{extended_metric}) may 
be seen as generalizations of PPN metrics, where the two Eddington parameters $\beta$ and $\gamma$ are replaced by two arbitrary functions  $\delta \Phi_N$ and $\delta \Phi_P$ of the 
radial distance to the gravitational source.

Assuming that the metric associated with the gravitational field of the Sun takes the form of a PPN metric,
predictions can be made for motions in this gravitational field and compared with observations \cite{Will06}.
Observations in the solar system provide constraints on the form of the single potential $\phi$ 
and on the  values of
 Eddington parameters $\beta$ and $\gamma$.
Modifications of Newton potential $\phi$  are usually parametrized 
in terms of an additional  Yukawa potential depending on two parameters, 
a range $\lambda$ and an amplitude $\alpha $ measured with respect to  $\phi$
\beqa
\label{Yukawa_perturbation}
\delta \phi(r) = \alpha e^{-\frac r\lambda} \phi(r)  
\eeqa
Corrections behaving as (\ref{Yukawa_perturbation}) have been looked for at various values of  $\lambda$
ranging from the millimeter scale \cite{Adelberger03} 
to the size of planetary orbits \cite{Coy03}.
For ranges of the order 
of the Earth-Moon \cite{Williams96} to Sun-Mars distances 
\cite{Hellings83,Reasenberg79,Kolosnitsyn03} tests have been performed by following the motions of planets and artificial probes. Although agreeing with GR, all these results still show \cite{Coy03,JR04} that windows remain open for violations of the Newton force law at short ranges $\lambda$, below the millimeter, as well as
long ones, of the order of or larger than the size of the solar system.
On another hand, experiments performed up to now have confirmed that, assuming that $\beta$ and $\gamma$ 
take constant values, the latter  should be close to $1$.
From Doppler ranging on Viking probes in the vicinity of Mars
\cite{Hellings83} to deflection measurements using VLBI astrometry \cite{Shapiro04} or radar ranging on the Cassini probe \cite{Bertotti03}, the allowed values for $\gamma-1$ have decreased with time.
Precessions of planet perihelion \cite{Talmadge88} and 
polarization by the Sun of the Moon orbit around the Earth 
\cite{LLR02}  constrain linear superpositions 
of $\beta$ and $\gamma$. As a result, in order to remain compatible with gravity tests, PPN metrics must satisfy
rather stringent constraints 
\beqa
\label{classical_tests}
|\gamma -1| \le 10^{-5}, \qquad |\beta -1| \le 10^{-4}
\eeqa

Obviously, gravity tests performed in the solar system provide  evidence for a metric theory lying very close to
GR. In case of a PPN metric (\ref{PPN_metric}), the latter should correspond to values of $\beta$ and $\gamma$ close to $1$.
These constraints however result from an analysis performed with the assumption of constant Edddington parameters, which only covers a small subspace of potential deviations from GR (\ref{generalized_PPN_metric}).
Dependences of $\beta$ and $\gamma$ on the length scale are not excluded and deviations at very short or very large scale are losely constrained. Gravity tests still leave room for alternative metric theories, such as metric extensions of GR  (\ref{extended_metric}), provided they satisfy these criteria. 
They should correspond to small anomalous potentials $\delta\Phi_N$ and $\delta \Phi_P$ so that to remain close to GR and thus compatible with present gravity tests.

As remarked in the introduction, deviations from GR may have already been observed in the very domain where classsical tests have been performed, namely within the solar system. The Pioneer 10/11 probes, which were launched in the seventies and tracked during their travel to the outer part of the solar system indeed showed anomalous behaviors.
The trajectories of the probes where determined from the radio frequency signals sent to them, transponded on board and received by stations on Earth. The Doppler shifts affecting the signals received on Earth, when compared with the emitted ones, delivered the velocity and hence, by integration,  the distance separating the probe from the Earth.
In fact, Doppler data provide the trajectory once  a modelization of all gravitation effects on light propagation and on geodesic trajectories has been made. It appeared that the modelization based on GR led to anomalies, taking the form of Doppler residuals,
{\it i.e.} differences between observed and modelized velocities $v_{obs} - v_{model}$,  which could not be reduced \cite{Anderson02a}. The latter furthermore took the form of a linear dependence in time $t$, {\it i.e.} of a  roughly constant acceleration $a_P$,
 over distances ranging from $20$ A.U to $70$  A.U., directed towards the Sun 
\beqa
\label{Pioneer_acceleration}
v_{obs} - v_{model} \simeq -a_P(t-t_{in}), \qquad a_P \simeq 0.8 \ {\rm{nm} \ \rm{s^{-2}}}
\eeqa
No satisfactory explanation in terms of systematic effects taking their origin on the probe itself or in its spatial environment has been found up to now \cite{Anderson07}. 
Recently, a larger set of data, covering  parts of the Pioneer 10/11 missions which had not been analysed,  has been recovered and put under scrutiny \cite{PAIT}. Several teams have engaged in an independent reanalysis of the Pioneer data  
\cite{Markwardt02,Olsen07,Levy08}. Proposals have also been made for  missions dedicated to a study of deep space gravity \cite{Dittus05}.
At present, the possibility that the Pioneer anomaly points at the necessity to change the theoretical framework cannot be ignored. In that case, the existence of an extended framework having the ability to account for the Pioneer anomaly while remaining consistent with all gravity tests constitutes a crucial element.

 PPN metrics which are compatible with classical gravity tests practically reduce to GR metric (see (\ref{classical_tests})).
Furthermore, the radial dependence of their curvatures (\ref{generalized_PPN_metric}) do not allow  PPN metrics to reproduce the properties of the  Pioneer anomaly (\ref{Pioneer_acceleration}).
On another hand, metric extensions of GR (\ref{GR_metric},\ref{extended_metric}) provide a phenomenological framework which enlarges the PPN neighborhood of GR, with corrections brought to curvatures which may remain significant at large distances from the gravitational source.  

When performed within the framework of  a general metric extension of GR, an explicit computation of the Doppler signal, in the physical configuration corresponding to the Earth and  a remote probe, exhibits a discrepancy with the similar signal computed using GR \cite{JR05b,JR06a}.
Computation is more easily achieved by using the time delay function, {\it i.e.} a two-point function which describes the elapsed time between emission and reception of a lightlike signal propagating between two spatial positions.
The time delay function is easily determined from the metric components by using a reference frame where the extended metric takes the form  of a stationary isotropic metric   \cite{JR06b}.
The Doppler signal corresponding to a tracking of the probes by stations on Earth is then obtained as the time derivative of  the time delay function evaluated on the  trajectories of the Earth and probes.
The time derivative of the Doppler signal itself may be written under the form of a time-dependent acceleration, as the observed Pioneer anomaly (\ref{Pioneer_acceleration})\cite{Anderson02a}. 
For an extended metric (\ref{extended_metric}) which is close to GR metric, dependences on metric components may be treated 
perturbatively and only the first order in metric perturbation may be kept.
Metric perturbations may be seen to modify the expression for the Doppler signal in two ways, through a perturbation of the time delay two-point function itself, and through a perturbation of the trajectories on which this function is evaluated.
A difference then emerges between the acceleration
obtained for an extended metric and that entailed by GR metric, which corresponds to the discrepancy which would appear when comparing an observed signal with the similar signal predicted using GR.
This difference can be considered as representing the anomaly which would be observed.

When expressed in terms of the anomalous potentials defining the extended metric (\ref{extended_metric}), the  anomalous acceleration contains several contributions due to different sources of perturbation, affecting in particular the  propagation of lightlike signals and the trajectories of massive bodies. We shall consider the simplified case of a remote probe which is moving on an escape trajectory in the ecliptic plane and which is reaching the outer part of the solar system. Several contributions to the anomalous acceleration may then be neglected and the latter takes a simplified expression
\cite{JR06b} (notation $[\quad]_{st}$ stands for expressions obtained using GR)
\beqa
\label{anomalous_acceleration}
&&\delta a \simeq \delta a_{sec} + \delta a_{ann}\nonumber\\
&&\delta a_{sec} \simeq -{c^2\over2} \partial_r(\delta g_{00})
+[\ddot{r}]_{st}\left\lbrace{\delta(g_{00}g_{rr})\over2} -\delta g_{00}\right\rbrace
-{c^2\over2}\partial_r^2[g_{00}]_{st}\delta r \nonumber\\
&&\delta a_{ann} \simeq {d\over dt}\left\lbrace[\dot{\phi}]_{st}\delta\rho\right\rbrace 
\eeqa
$(r, \phi)$ represents the position of the probe with respect to the Sun, in the ecliptic plane ($r$ is the radial distance and $\phi$ the difference of angular positions between the probe and the Earth) and  $\rho$ denotes the impact parameter (with respect to the Sun) of the  lightlike signal propagating  between the probe  and the Earth. The acceleration anomaly divides into two parts, a secular anomaly $\delta a_{sec}$ which varies over large times only, typically for variations of the distance between Earth and probe of the order of several A.U., and a modulated  anomaly 
 $\delta a_{ann}$ which describes variations with annual or semi-annual periodicity.
One notes that, due to several simplifications in the previous representation (neglecting in particular the motions of the stations, effects of the Earth atmosphere, deviations from the ecliptic plane, etc ...) additional modulations, in particular with daily periodicity, have been ignored. 
Both parts of the anomalous acceleration are generated by anomalies in the two gravitational potentials ($\delta g_{00},  \delta g_{rr}$) and in the probe trajectory ($\delta r$). In particular, the latter cannot be ignored in the case of  Pioneer 10/11 probes as no range capabilities (allowing one to directly determine the radial position of the probe) were available for them.  
Anomalous metrics  (\ref{anomalous_acceleration}) can be seen to lead to anomalous accelerations which exhibit the same  qualitative features as the Pioneer anomaly \cite{Anderson02a,Markwardt02,Olsen07,Levy08}. 

To obtain from the observed Pioneer anomaly quantitative constraints on the two gravitational potentials, one needs to enter a  detailed analysis of navigation data, performing this analysis in the enlarged  phenomenological
framework provided by metric extensions of GR.  
As already shown in the above simplified model (\ref{anomalous_acceleration}), the acceleration anomaly exhibits a secular  part and modulations which both depend on anomalies in the gravitational potentials and in the trajectory.
Correlated anomalies  should then play an essential role when confronting models pertaining to the extended phenomenological framework with a detailed analysis of Pioneer data \cite{JR06b}. 

Although a precise confrontation is not yet available, some consequences can nonetheless be drawn from the general form of the secular anomaly observed on the Pioneer pobes.
As shown by equations (\ref{anomalous_acceleration}) (anomaly modulations are used to eliminate the trajectory anomaly  \cite{JR06b}), a constant anomalous acceleration $\delta a = -a_P \equiv -{c^2\over l_H}$ over the distances covered by the Pioneer probes may follow from different forms of the anomalous gravitational potentials (\ref{extended_metric}).
If only a perturbation of Newton potential is assumed, the latter should behave linearly with the heliocentric distance,
{\it i.e.} $\delta \Phi_N \simeq r/l_H$, to produce a constant acceleration.  This form however conflicts with tests which have been performed  between the Earth and Mars \cite{Reasenberg79,Anderson02a,JR05a}, so that an  anomaly limited to  Newtonian sector is only allowed at large heliocentric distances \cite{Moffat06}, where it must remain compatible with the ephemeris of outer planets. On another hand, a constant acceleration is also obtained for an anomaly in the second sector which behaves quadratically with the heliocentric distance, {\it i.e.} $\delta \Phi_P \simeq  {c^2\over 3 G_N M}{r^2\over l_H}$, and a combination of anomalies in the two sectors may also lead to the form taken by Pioneer data \cite{JR06b}.
One notes that these properties cannot be obtained from  PPN metrics (\ref{generalized_PPN_metric}) and that the form  just given for the anomalous potential in the second sector  $\delta \Phi_P$ corresponds to a non vanishing constant curvature in the outer part of the solar system.

A remarkable feature of the phenomenological framework derived from metric extensions of GR (\ref{GR_metric},\ref{extended_metric}) is that, besides producing Pioneer-like anomalies, it also allows one to 
preserve the agreement with classical gravity tests.
An important part of gravity tests is provided by ephemerids of inner planets, and in particular by perihelion precession anomalies \cite{Hellings83}. Anomalies in perihelion precessions of planets (with respect to Newtonian gravitation) are well accounted for by the non linear dependence of GR metric on Newton potential (\ref{PPN_metric}). Metric extensions (\ref{GR_metric},\ref{extended_metric}) induce modifications which generalize the corrections obtained in the PPN framework (\ref{generalized_PPN_metric}). The anomalous gravitational potentials $\delta \Phi_N$ and $\delta \Phi_P$ may be seen as promoting the parameters $\beta$ and $\gamma$ to functions which depend on the heliocentric distance  \cite{JR06a}. Observed anomalies in perihelion precessions  constrain a particular combination of the two gravitational potentials for ranges around a few A.U. where planet ephemerids are known with precision. Provided the latter combination remains small within these ranges, compatibility of extended metrics with gravity tests in ensured.

Precise gravity tests are obtained by measurements of the deflection induced on a lightlike signal by the Sun gravitational field, as for instance those performed with Cassini probe \cite{Bertotti03}. Light deflection measurements can be seen to explore the behavior of gravitational potentials in the Sun vicinity, {\it i.e.} at small heliocentric distances. The deflection angle is determined by the impact parameter of the lightlike signal and, within a PPN framework,  is directly related to the value of the parameter $\gamma$. In the  framework of metric extensions  (\ref{GR_metric},\ref{extended_metric}),  the effect of anomalous gravitational potentials on the deflection angle may  be interpreted as due to a generalized  parameter $\gamma$, with a functional dependence on the impact parameter \cite{JR05a}. Deviations of this  generalized parameter  from its constant value corresponding to GR 
are given by a combination of the two anomalous gravitational potentials  (\ref{extended_metric}), so that light deflection measurements provide constraints on the behavior of gravitational potentials in the range of a few solar radii. 
A precise analysis shows  that these constraints still allow different behaviors of the two gravitational potentials with respect to the radial distance \cite{JR05b}. When compared with the value predicted by GR or even a PPN metric, the  deflection angle  might then show an anomalous behavior which becomes observable for large values of the impact parameter. Although  light deflection is more important,
 within the PPN framework, for small impact prameters and is thus usually measured with lightlike signals grazzing the Sun surface, 
 a global mapping of light deflection over the sky, including large impact parameters, 
is programmed in the future mission GAIA \cite{Gaia}. A detected anomalous behavior of deflection angles could point at the necessity
to use an enlarged phenomenological framework like that  provided by metric extensions of GR.

\section{Conclusion}

Despite difficulties which emerge at their interface, quantum theory and relativity theory cannot be considered
as independent frameworks, with quantum applications  excluding motion and gravitation and  relativity being limited
to a representation of classical systems. Quantum fields possess fluctuations which result in mechanical effects and 
affect inertial responses. This entails a revision of our notions of mass and motion which becomes necessary when dealing with microscopic systems. Quantum field fluctuations also modify gravitation with the possibility of leading to observable effects at macroscopic scales. The effects of quantum fluctuations on relativistic systems cannot be ignored 
and it appears necessary to have a representation which is compatible with both frameworks.

We have shown that quantum field fluctuations, even in vacuum,  
must be taken into account when analysing mechanical effects such as inertia.
As exemplified on Casimir energy and contrarily to a common opinion, vacuum field fluctuations do contribute to inertia and the induced inertial mass complies  both with quantum and with relativistic requirements.
These properties are more easily put into evidence using the linear response formalism, which thus allows one to connect in a consistent way the quantum and relativistic frameworks.  Infinities which are responsible for difficulties when interfacing  quantum and relativity theories can be treated by renormalization techniques. The linear response formalism not only recovers the fundamental connection between interacting and free quantum fields lying at the basis of QFT, but also allows one to implement in a general way the fundamental relation made by relativity between motions and the symmetries of space-time. 
A treatment of motion in quantum vacuum may thus be given which remains  consistent with the relativistic conception of space-time
and which brings new light on the role played by quantum vacuum with respect to inertia \cite{JR97b,JR02}.

The effects  of quantum fluctuations on inertia have important consequences for the notion of mass.
There results that the mass parameter used for characterizing motion is a quasistatic approximation. Mass possesses quantum fluctuations which cannot be neglected in high frequency regimes, or equivalently at very short time scales, and these fluctuations should be described by the correlations of a quantum operator representing the mass observable \cite{JR93d}.
This modification of the representation of mass is accompanied by similar extensions for positions and motions in space-time in order to comply with the requirements imposed on observables by quantum and relativity theories. 
Remarkably, these extensions are ensured in a consistent way by the existence of a single algebra where quantum observables and relativistic transformations are simultanously represented. This unique algebra, built on space-time symmetries, provides 
relations between mass and motions, including uniformly accelerated motions, which hold in the quantum framework.
This allows in particular to write a quantum version of Einstein effect and  shows that the equivalence principle can be extended to the
quantum framework \cite{JR97a}.

According to the principles of relativity, quantum field fluctuations affect not only inertial but also gravitational masses, with the consequence that classical gravitation must be modified to account for radiative corrections.
Presently, theoretical analysis and experimental observations both support the equivalence principle as
 a faithfull and well tested basis for a gravitation theory. These arguments strongly favor  metric theories but 
do not constrain the gravitational coupling. 
If classical tests performed up to now in the solar system tend to confirm 
Einstein-Hilbert Lagrangian, the corresponding coupling looses its simple form when 
corrections induced by quantum fluctuations are taken into account.
Although applying renormalization techniques to gravitation theory appears as a formidable task, 
theoretical arguments and observations both point at the possibility of corrections to gravitation occuring at the classical level, {\it i.e.} at large length scales.
Radiative corrections induce modifications which amount to replace Newton gravitation constant by two gravitation running coupling constants corresponding to two sectors with different conformal weight. These may equivalently be represented by metric extensions 
of GR which can be characterized either by two non vanishing Ricci curvatures or by two gravitational potentials. 
Besides modifications of Newton potential,
a gravitational potential in a second sector opens new phenomenological possibilities.  
Metric extensions of GR appear as an efficient way to parametrize gravitation theories lying in the neighborhood of GR 
and offer larger possibilities than the usual PPN phenomenological framework. Having  also the 
ability to remain compatible with classical gravity tests, 
they  may  account for Pioneer-like anomalies and may further lead to
other correlated anomalies which could be observed in future experiments \cite{JR07a}.

\end{document}